\documentclass[showkeys,nofootinbib,prd]{revtex4}

\usepackage{amsmath}
\usepackage{amsfonts}
\usepackage{amssymb}
\usepackage{amsthm}
\usepackage{mathtools}
\usepackage{subfigure}
\DeclareFontFamily{U}{mathb}{\hyphenchar\font45}
\DeclareFontShape{U}{mathb}{m}{n}{
      <5> <6> <7> <8> <9> <10> gen * mathb
      <10.95> mathb10 <12> <14.4> <17.28> <20.74> <24.88> mathb12
      }{}
\DeclareSymbolFont{mathb}{U}{mathb}{m}{n}

\DeclareMathSymbol{\Sun}{3}{mathb}{"40}
\allowdisplaybreaks

\begin{document}
\title{Cosmological extra dimensions can mimic dark matter I. Dark matter in galaxies}
\author{Mattia Villani}
\affiliation{University of Urbino Carlo Bo, Department of Pure and Applied Sciences (DiSPeA), Via Santa Chiara, 27, Urbino (PU), 61029, Italy}
%\affiliation{Independent scholar}
\email{mattia.villani@uniurb.it}
%\email{mattiav25@gmail.com}

\begin{abstract}
We present a simple model with a $n+4$-dimensional metric with compactified extra dimensions with a space dependent radius $R$. We show that changes in the profile of $R$ are sourced by (visible) matter density and Newtonian potential. We can reproduce rotation curves of spiral galaxies and provide an expression for $R$. In order to test our hypothesis, we propose six experiments.
\end{abstract}
\maketitle

\section{Introduction}

The presence of a dark form of matter was hypothesized in the 1930s due to the works of Oort and Zwicky \cite{oort,zwicky}. Later, the work of Vera Rubin has shown that the rotation curve of spiral galaxies cannot be explained by considering only the luminous matter and gas present in the galaxies. Modern cosmological observations with Supernovae \cite{snae} or satellites for the study of the CMB like Planck \cite{planck} have shown that the so-called baryonic mass is only about 5\% of the total mass of the Universe; there is also a much larger contribution due to dark matter, which brings the total contribution of matter to about 31\% and a 69\% contribution due to dark energy, a form of energy that is leading the current accelerated expansion of the Universe.

There are several theories on the nature of dark matter; we know that it should be composed of particles that do not interact with electromagnetic fields, but only with gravitational field and possibly weak interactions and that it must be cold, non-relativistic, to explain the structure formation in the Universe. One popular proposal for dark matter is that it is composed of the so-called WIMPs (weakly interacting massive particles, \cite{wimp}), supersymmetric particles, such as the neutralino, which are stable and neutral. Another options are axions, very light particles introduced to solve a problem in QCD \cite{weinberg}, but also popular in String Theory \cite{axion}. Sterile neutrinos have also been considered \cite{ster}. Finally, it is possible that dark matter is composed of primordial black holes \cite{pbh}. None of the above hypotheses has been confirmed \cite{search} and the nature of dark matter is still a mystery.

In several papers \cite{RS1,RS2,RS3,RS4,RS5,RS6}, authors considered a metric of the form
\begin{equation}
    ds^2=dt^2-\dfrac{a(t)^2}{1-r^2\,\kappa}dr^2-a(t)^2\,r^2\,d\Omega_2-R(t)^2\,d\Omega_n
\end{equation}
i.e. a $n+4$-dimensional metric in which the 4d submanifold is a FLRW spacetime, while the extra dimensions are compactified to a hypersphere with time-dependent radius $R$ in principle different from the scale factor $a$ of the 4d spacetime. Here we focus on the scale of galaxies and galaxy clusters considering a metric of the form
\begin{equation}\label{eq:met}
    ds^2=-(1+2\Phi(r))\,dt^2+dr^2+r^2\,d\Omega_2-R(r)^2\,d\Omega_n
\end{equation}
a $n+4$-dimensional metric in which the 4d submanifold is a perturbed Minkowski metric in the presence of a Newtonian potential $\Phi\ll1$, assumed small and a $d$-dimensional hypersphere with a space-dependent radius $R$, i.e. we assume that at small scale the radius of the extra dimensions is essentially constant in time, but it depends on the radial coordinate $r$. The aim of the work is to prove that $R$ behaves as an additional potential that we observationally associate to the presence of dark matter. The need for a (fake) dark matter halo that surrounds galaxies is therefore a consequence of the failure to detect the extra dimensions. We also provide six possible experiments (not new, but that can be performed with present day technology) that can verify our hypothesis: the study of the velocity of galaxy satellites, the study of elliptical and spiral galaxies, the study of the motion of the S2 star around Sgr A$^{*}$ black hole, the study of the shadow of this black hole or M87$^{*}$ and the study of time delays. In all these experiments, the extra dimensions give a contribution, and using the explicit form for $R(r)$ that we derive in our work, one can verify their presence.

This paper is organized as follows: in Section \ref{sec:potential}, we introduce the effective potential containing the effect of the radius $R$; in Section \ref{sec:rot}, we prove that we can reproduce the rotation curve of spiral galaxies; in Section \ref{sec:discussion} we discuss our result and present the five experiments we propose; in Section \ref{sec:concl} we conclude our exposition. There are two appendices: in Appendix \ref{sec:ein}, we derive the Einstein equations, while in Appendix \ref{sec:geod}, we derive the geodesics of a higher dimensional Kerr metric.

\section{The effective potential and dark matter}
\label{sec:potential}

For the moment, we focus on a 6-dimensional spacetime, but the results are in fact independent on the number of dimensions. From the metric \eqref{eq:met}, we can derive the geodesics equations using the Hamilton-Jacobi method. The Hamilton-Jacobi equation for the 6 dimensional case is
\begin{equation}
    -\dfrac{dS}{d\lambda}=-(1-2\Phi)\,\dfrac{1}{2}\,\left(\dfrac{dS}{dt}\right)^2+\dfrac{1}{2}\,\left(\dfrac{dS}{dr}\right)^2+\dfrac{1}{2r^2}\,\left( \dfrac{dS}{d\theta} \right)^2+\dfrac{1}{2r^2\,\sin^2\theta}\,\left( \dfrac{dS}{d\phi} \right)^2+\dfrac{1}{2R^2}\,\left[ \left( \dfrac{dS}{d\theta_1} \right)^2 + \dfrac{1}{\sin^2\theta_1}\,\left( \dfrac{dS}{d\theta_2} \right)^2 \right].
\end{equation}
We make the ansatz 
\begin{equation}
    S=\dfrac{m^2}{2}-E\,t+L_z\,\phi+L_1\,\theta_2+S_r(r)+S_\theta(\theta)+S_{\theta_1}(\theta_1),
\end{equation}
where $m$ is the mass of the particle, $E$ its energy, $L_z$ is the third component of the angular momentum and $L_1$ a sort of angular momentum relative to the extra dimensions. Then we find 
\begin{equation}
    -m^2\,r^2\,R^2=-(1-2\Phi)\,E^2\,r^2\,R^2+r^2\,R^2\left( \dfrac{dS_r}{dr} \right)^2+R^2\,\left[ \left( \dfrac{dS_\theta}{d\theta} \right)^2+\dfrac{L_z^2}{\sin^2\theta} \right]+ r^2\,\left[\left( \dfrac{dS_{\theta_1}}{d\theta_1} \right)^2 + \dfrac{L_1^2}{\sin^2\theta_1}\right]
\end{equation}
by separation of variables, we obtain
\begin{align}
    \left(\dfrac{dS_r}{dr}\right)^2&=-m^2-(1-2\Phi)\,E^2+\dfrac{L^2}{r^2}+\dfrac{L_2^2}{R^2}=\mathcal{R},\\
    \left(\dfrac{dS_\theta}{d\theta}\right)^2&=L^2-\dfrac{L_z^2}{\sin^2\theta}=\Theta,\\
    \left(\dfrac{dS_{\theta_1}}{d\theta_1}\right)^2&=L_2^2-\dfrac{L_1^2}{\sin^2\theta}=\Theta_1
\end{align}
where $L$ and $L_2$ are separation constants (of course $L$ can be interpreted as the total angular momentum). Now we derive the action in $m^2$, $L_z$, $L_1$, $L$, $L_2$ and $E$, obtaining respectively,
\begin{align}
    \lambda&=\int\dfrac{dr}{\sqrt{\mathcal{R}}},\\
    \phi&=\int \dfrac{L_z\,d\theta}{\sin^2\theta\,\sqrt{\Theta}},\\
    \theta_2&=\int \dfrac{L_1\,d\theta_1}{\sin^2\theta_1\,\sqrt{\Theta_!}},\\
    0&=\int \dfrac{dr}{r^2\,\sqrt{\mathcal{R}}}+\int\dfrac{d\theta}{\sqrt{\Theta}},\\
    0&=\int\dfrac{d\theta_1}{\sqrt{\Theta_1}},\\
    t&=\int\dfrac{(1-2\Phi)\,E\,dr}{\sqrt{\mathcal{R}}}.
\end{align}
Combining the above equations, we can obtain the geodesics, in particular, we find
\begin{align}
    \dfrac{dr}{d\lambda}&=\sqrt{\mathcal{R}}=\sqrt{-m^2-(1-2\Phi)\,E^2+\dfrac{L^2}{r^2}+\dfrac{L_2^2}{R^2}},\\
    \dfrac{dt}{d\lambda}&=(1-2\Phi)\,E.
\end{align}
We can identify in the above expressions an effective potential
\begin{equation}\label{eq:eff_pot}
    \Phi_{eff}=\Phi+\dfrac{L_2^2}{R^2}.
\end{equation}
The first is the Newtonian potential due to the luminous mass, and we claim that the second is what we call dark matter, which in this case is a pure extra dimension effect.

\section{Spiral galaxies rotation curves}
\label{sec:rot}
It is known that one major observational clue of the presence of a  dark form of matter is the fact that rotation curves of spiral galaxies do not fall off as $r^{-1/2}$ at infinity, but remain constant (or even grow in some cases). This can be explained by introducing an halo of dark matter that contains all the visible forms of matter. A density profile often used to describe this halo is the Navarro-Frenk-White (NFW) profile \cite{nfw,nfw2}
\begin{equation}
    \rho_{DM}(r)=\dfrac{\rho_0\,a^3}{r\,(a+r)^2},
\end{equation}
where $\rho_0$ is the central density and $a$ is a scale length.

Given the (radius-dependent) rotation velocity $V(r)$, using simple Newtonian physics, we can obtain the following expression that relates the velocity to our effective potential
\begin{equation}\label{eq:profile}
    \dfrac{V(r)^2}{r}=-\dfrac{d\,\Phi_{eff}}{dr}=-\dfrac{d}{dr}\,\left( \Phi+\dfrac{L_2^2}{R(r)^2} \right)
\end{equation}
The velocity will have two contributions $V=V_v+V_d$, where $V_v$ will be given by the usual Newtonian potential, while $V_d$ will be given by the second term and in the usual interpretation is the contribution due to dark matter.

Using again plain Newtonian physics, we find that the mass contained in a sphere of radius $r$, $M(<r)$ is given by
\begin{equation}\label{eq:vel_mass}
    M(<r)=\dfrac{V(r)^2r}{G}.
\end{equation}
The mass is also obtained integrating the density profile
\begin{equation}
    M(<r)=\int dr\,4\pi\, r^2 \rho(r)
\end{equation}
If we focus only on  the (fake) \emph{dark} component and use the NFW profile for the fake dark matter halo, we obtain
\begin{equation}\label{eq:mass}
    M(<r)=4\pi \rho_0\,a^3\,\left[ \dfrac{a}{a+r}+\ln\left( \dfrac{a+r}{a} \right) \right].
\end{equation}
Then combining \eqref{eq:vel_mass} and \eqref{eq:mass}, we find an expression for $V_d$. Finally, equation \eqref{eq:profile}, neglecting the visible contribution, gives a differential equation for $R(r)$. Integrating and imposing that $R(r\rightarrow\infty)=r_0$, we find
\begin{equation}\label{eq:def_R}
    R(r)=\dfrac{L_2}{\sqrt{\dfrac{L_2^2}{r_0^2}+\dfrac{4\pi G \rho_0}{r}\,\left( 1+\ln\left(\dfrac{a+r}{a} \right) \right)}}
\end{equation}
The limit for small $r$ is
\begin{equation}\label{eq:small}
    R(r\rightarrow0)=\dfrac{L_2}{2}\,\sqrt{\dfrac{r}{\pi\,G \, \rho_0}}
\end{equation}

The above calculations show that it is possible to describe the rotation curves of the spiral galaxies in a higher dimensional spacetime using equation \eqref{eq:def_R} for the expression of the extra dimensions radius. It depends on three parameters: the value of $R$ at infinity $r_0$, which is the cosmological value of $R$ obtained solving the Friedman equations reported in \cite{RS1,RS2,RS3,RS4,RS5,RS6}, $\rho_0$, the central density of the fake dark matter halo and $a$, its scale length. Other choices of the fake dark matter profile will give different expressions for $R$, however, we notice that the chosen one is well-behaved both at large and small radii.

\section{Discussion}
\label{sec:discussion}

In the above Sections, we have found that the radius of the extra dimensions behaves as an additional potential and we have found an expression for $R$ which can reproduce the rotation curve of spiral galaxies without the need to introduce a dark matter component. We need to answer some question:
\begin{enumerate}
    \item Where does the change in the profile of the radius $R$ come from?
    \item Some galaxies are known to not have dark matter \cite{NoDM}: how can we explain this phenomenon with our approach?
    \item Other galaxies are, instead, composed mostly of dark matter \cite{dark,dark2,dark3,dark4}: can we explain this?
    \item Can we describe the mass distribution of the Bullet Cluster?
    %\item Can we explain the Bullet Cluster?
\end{enumerate}

\begin{figure}
    \centering
    \subfigure{\includegraphics[width=0.5\linewidth]{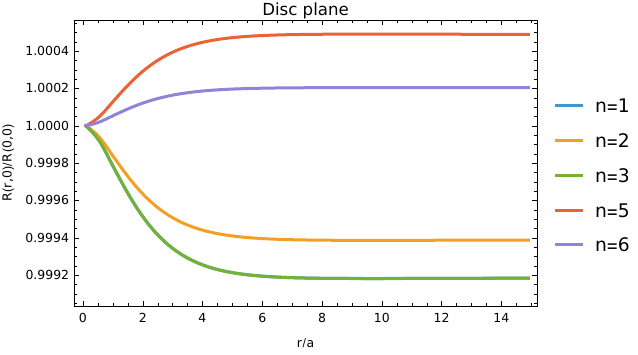}}
    \subfigure{\includegraphics[width=0.5\linewidth]{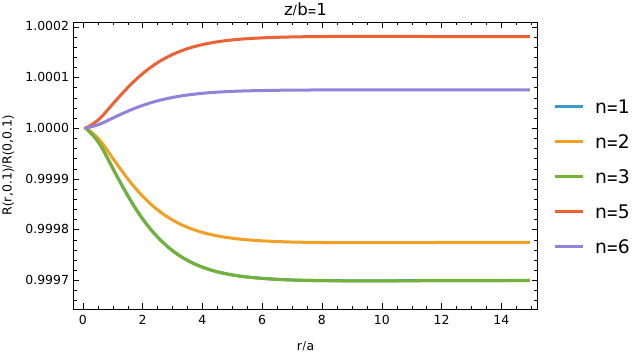}}
    \subfigure{\includegraphics[width=0.5\linewidth]{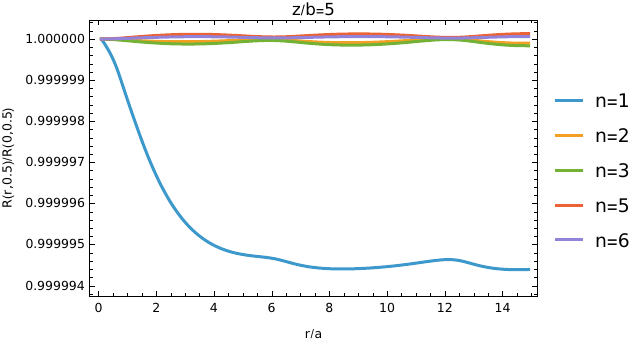}}
    \subfigure{\includegraphics[width=0.5\linewidth]{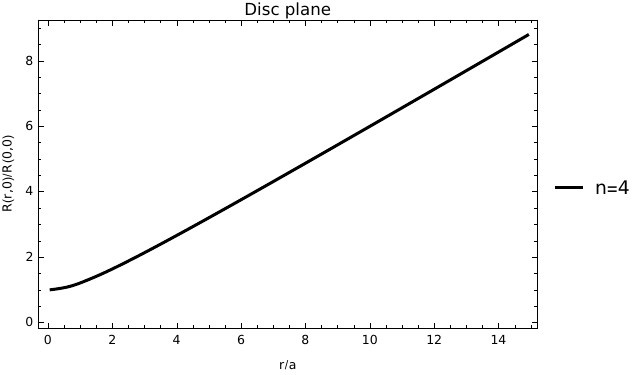}}
    \caption{The evolution of $R$ as a function of the radial coordinate $r$ for three different position on the $z$ axis. Top: disc plane ($z=0$); Second panel: $z=b$; Third panel: $z=5b$, where $b$ is the scale length of the thickness of the disc; Bottom panel: the case $n=4$ on the disc plane. These plots where obtained imposing $M=10^{10}$ M$_\Sun$, $a=1$ kpc $b=100$ pc.}
    \label{fig:doubleexp}
\end{figure}

We can answer the first question by looking at the Einstein equations of the metric \eqref{eq:met}. They can be reduced to a single equation (see the Appendix \ref{sec:ein})
\begin{equation}\label{eq:final}
    \dfrac{R^\prime}{R}=\dfrac{8\pi\,G\,\rho\,(1-2\Phi)-2n\,\Phi^\prime-2r\,\Phi^{\prime\prime}}{n(n-4)}, \qquad R=k \, \exp\left[ \int dr \left( \dfrac{8\pi\,G\,\rho\,(1-2\Phi)-2n\,\Phi^\prime-2r\,\Phi^{\prime\prime}}{n(n-4)} \right) \right].
\end{equation}
This equation is not valid for $n=4$, in that case one should use
\begin{equation}\label{eq:final2}
    \left( \dfrac{R^\prime}{R} \right)^2+2\,\dfrac{R^\prime}{R}-\dfrac{1}{R^2}=-\dfrac{8\pi G\,\rho}{12}\,\left( 1-2\Phi \right)-\dfrac{\Phi^\prime}{2r}.
\end{equation}
These equations show that any modification of the profile of $R$ is sourced by matter density and the Newtonian potential it generates. As we have seen, $R$ is correlated to the fake dark matter potential, thus, in this framework, visible matter is the real source of what we call dark matter. As an example on the usage of equation \eqref{eq:final}, we use the Mestel disc potential \cite{mestel}
\begin{equation}\label{eq:mestel}
    \Phi=v^2\,\ln\left[ \sqrt{r^2+z^2}+|z| \right]
\end{equation}
where $v$ is the circular velocity in the mid-plane and $z$ the thickness of the disc. Substituting into \eqref{eq:final}, we obtain
\begin{equation}
    R=k\,\left( \dfrac{2z}{z+\sqrt{r^2+z^2}} \right)^{\frac{2v^2\,(n-1)}{c^2\,n\,(n-4)}}\,\exp\left( -\dfrac{2v^2}{c^2}\,\dfrac{1-2r-\dfrac{z}{\sqrt{r^2+z^2}}}{n\,(n-4)} \right),
\end{equation}
where $k$ is an integration constant. $R$ has the asymptotic expansion
\begin{equation}
    R(r\rightarrow\infty)=k\,\left( \dfrac{2z}{r} \right)^{\frac{2v^2\,(n-1)}{c^2\,n\,(n-4)}}\,\exp\left( -\dfrac{2v^2}{c^2}\,\dfrac{r-2r^2+(n-2)\,z}{n\,(n-4)\,r} \right)
\end{equation}
We see that $R\rightarrow0$ as $r\rightarrow\infty$ for $n<4$ and $R\rightarrow\infty$ for $n>4$, while in any case $R\rightarrow k$ as $r\rightarrow0$, so \eqref{eq:mestel} might not be the best choice for the potential, since we expect that $R$ tends to a very small but finite value at infinity, however, it is simple enough to have an analytical integral. A better choice for a density profile could be a double exponential of the form
\begin{equation}
    \rho(r,z)=\dfrac{M}{2\pi\,a^2\,b}\,\exp\left( -\dfrac{r}{a} \right)\,\exp\left( -\dfrac{z}{b} \right)
\end{equation}
where $M$ is the (baryonic) mass of the disc and $a$ and $b$ are scale lengths. Using this profile, we obtain (numerically) the plots reported in Figure \ref{fig:doubleexp}. We notice that $R$ decreases for $n\leq3$ and grows for $n\geq5$, but in both cases reaches asymptotically a constant value. Moreover, we notice that the farther we go from the disc plane, the smaller the variation of $R$ and that at 5 scale length in thickness, $R$ reaches the asymptotic value. In the case of $n=4$ the solution derived from equation \eqref{eq:final2} seems to grow indefinitely.

In this approach, we can explain the fact that there are galaxies without dark matter: their density (and, therefore, their potential) is too small to source relevant modification of the profile of $R$ which could be interpreted as dark matter. In Table \ref{tab:density}, we report, as an example, the surface density of four galaxies (NGC 1052-DF2 \cite{DF2,DF2b}, NGC 1052-DF4 \cite{DF4}, NGC 1052-DF9 \cite{DF9}, FCC 224 \cite{FC224}) known to be without dark matter compared to the surface density of the Milky Way and of the Large Magellanic Cloud (LMC), which, instead, do have dark matter: we see that the former galaxies have a density several times smaller than that of the latter. We have no information on their Newtonian potential, however it is plausible that it is also smaller than that of the Milky Way and of the LMC, thus our hypothesis seems to be confirmed.

\begin{table}[htb]
    \centering
    \begin{tabular}{c|ccc}
        Galaxy & Luminous mass ($M_\Sun$) & Radius (kpc) &  Surface density ($M_\Sun\,\text{pc}^{-2}$) \\
        \hline
         NGC 1052-DF2          & $2\times10^{8}$   & 2.34 & 3.0\\
         NGC 1052-DF4          & $1.5\times10^8$   & 1.5 &  4.5\\
         NGC 1052-DF9          & $1.4\times10^8$   & 2.2 &  2.3 \\
         FCC 224               & $1.7\times10^8$   & 1.89 & 4.5\\
         \hline
         Milky Way             & $6.43\times 10^{10}$ & 13 &  30.3\\
         Large Magellanic Cloud& $10^{10}$         & 5  &  31.8
    \end{tabular}
    \caption{Surface density of some galaxies. In the first column we report the name of the galaxy, in the second the mass, in the third the radius and in the fourth the surface density. The first three rows report data from galaxies without dark matter, the other two report example galaxies known for the presence of dark matter. Data on radius and mass taken from \cite{DF2,DF4,DF9,FC224,MW,LMC}}
    \label{tab:density}
\end{table}

Dark galaxies are a particular type of galaxy composed at least of 99\% of dark matter; the baryonic mass is essentially gas and dust, with a small contribution due to stars. There are several candidates such as CDG-2 \cite{dark3} and Dragonfly 44 \cite{dark4}; these two galaxies have a very small luminous mass $\approx 10^7$ M$_\Sun$, but a relatively large radius $\approx 1.5$ kpc, making them ultra diffuse. We could not find a way to explain this type of galaxies within our framework, in fact they have a very low density $\approx 0.4$ M$_\Sun$ pc$^{-2}$, thus they should behave as the galaxies without dark matter discussed above. By inspecting equation \eqref{eq:final}, we can conclude that there must be something special in their Newtonian potential, making the derivative large $\Phi^\prime,\Phi^{\prime\prime}\gg1$. This requires further studies. 

Similarly, we need further studies on the Bullet cluster. The Bullet Cluster is a galaxy cluster in which the distribution of luminous matter does not coincide with the distribution of \emph{dark} matter and it is often quoted as another proof of the existence of the latter \cite{bullet}. As we shall see in a forthcoming paper, gravitational lensing is also affected by the presence of the extra dimensions, and the mass reconstruction procedure should be modified accordingly; thus, it is possible that the Bullet Cluster might be explained in this framework, but we need further studies.

In order to verify our hypothesis on the presence of extra dimensions and on the effects and profile of $R$ we propose six experiments, three of which can be performed inside the Galaxy, while three require extra galactic observations:
\begin{enumerate}
    \item Study of the velocity distribution of satellite galaxies;
    \item Study of elliptical galaxies;
    \item Study of spiral galaxies rotation curve;
    \item Study of the motion of S stars around the Sgr A$^{*}$ black hole;
    \item Study of the shadow of the Sgr A$^{*}$ black hole;
    \item Study of time delays.
\end{enumerate}
We analyze the five proposals in turn.

\subsection{Velocity distribution of satellite galaxies}

Numerous authors have used kinematical properties of satellite galaxies to trace the profile of dark matter halo, for an incomplete list see \cite{sat,sat2,sat3,sat4,sat5} and references therein. In our case, the radial velocity dispersion $\sigma^2$ is given by the differential equation \cite{sat6}
\begin{equation}
    \dfrac{d\sigma^2}{dr}+\dfrac{\sigma^2}{r}\,\left( \beta + 2 \dfrac{dln\rho}{d\ln r} \right)=-\Phi_{eff}^\prime=-\Phi_N^\prime-\dfrac{d}{dr}\left( \dfrac{L_2^2}{R^2} \right), \qquad \beta=1-\dfrac{\sigma_\perp^2}{2\sigma^2}
\end{equation}
A formal solution is given by
\begin{equation}
    \sigma^2=\dfrac{1}{\chi\,\rho}\,\int dr\,\chi\,\rho\, \Phi^\prime_{eff}, \qquad \chi=\exp\left(\int dr\,\dfrac{2\beta}{r}\right).
\end{equation}
Following \cite{sat6}, one needs an explicit expression for $\rho$, $\beta$ and $\chi$. $\Phi_N$ is given by the luminous mass potential, while $R$ is given by equation \eqref{eq:def_R} for spiral galaxies. Comparing the resulting $\sigma^2$ with the observed one, one could verify our hypothesis that dark matter is an effect of the extra dimensions and also fix $r_0$. We could get a constraint on $n$ if instead of considering equation \eqref{eq:def_R}, we use a numerical approach similar to the discussion above about the double exponential disc, i.e. one could derive theoretically the velocity given a (baryonic) matter profile from equations \eqref{eq:eff_pot} and \eqref{eq:final} or \eqref{eq:final2} for different values of $n$ and compare the results to the observations.

\subsection{Elliptical galaxies}

\begin{figure}
    \centering
    \includegraphics[width=0.6\linewidth]{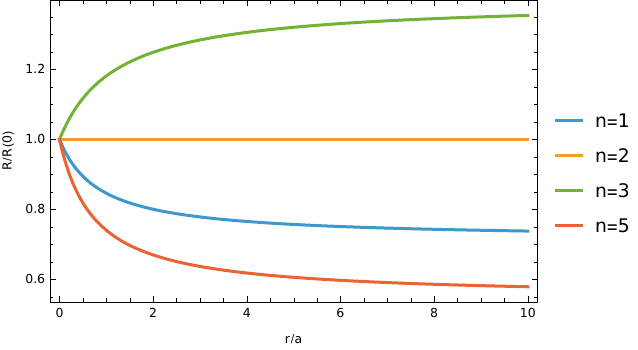}
    \caption{The evolution of $R$ as a function of $r/a$ for a Hernquist potential as a function of $n$, normalized to the value in $r=0$.}
    \label{fig:hern}
\end{figure}

The potential of elliptical galaxies is frequently described by the Hernquist profile \cite{hernquist}
\begin{equation}\label{eq:hern}
    \Phi=-\dfrac{1}{c^2}\dfrac{G \, M}{r+a}, \qquad \rho=\dfrac{M}{2\pi}\,\dfrac{a}{r\,(a+r)^3}
\end{equation}
where $M$ is the mass of the galaxy and $a$ is a scale length. This profile is sufficiently simple so that equation \eqref{eq:final} can be integrated analytically;\footnote{We could not find a solution to equation \eqref{eq:final2} with this profile.} the result is
\begin{equation}\label{eq:hern2}
    R=k\,\exp\left( -\dfrac{GM\,(n-2)\,r}{a\,c^2\,n\,(n-4)\,(a+r)} \right),
\end{equation}
where $k$ is an integration constant. We obtain the asymptotic expansion
\begin{equation}\label{eq:asy}
    R\rightarrow r_0=k\,\exp\left( -\dfrac{GM\,(n-2)}{a\,c^2\,n\,(n-4)} \right)
\end{equation}
These expressions can be used together with the Newtonian potential \eqref{eq:hern} to predict the velocity dispersion in an elliptical galaxy and compare the result with observations: this will fix the constants $a,k$, estimate $n$ and thanks to equation \eqref{eq:asy} we can obtain the asymptotic value $r_0$. We report in Figure \ref{fig:hern} the schematic behavior of equation \eqref{eq:hern2} for different values of $n$ (the mass $M$ was included within the integration constant, thus it simplifies when taking the ratio $R/R(0)$): we see that $R$ is constant for $n=2$, grows for $n=3$ and decreases otherwise. We also notice that $R$ can vary quite substantially with $r$: for $n=1$, for example, the asymptotic value is only 80\% of the value for $r=0$, for $n=5$ it is the 60\%.

\subsection{Spiral galaxies rotation curve}
Similarly to the above experiment, the study of spiral galaxies rotation curve can be used to constrain the value of $R$. In this kind of experiment, one should use equations \eqref{eq:profile} and \eqref{eq:eff_pot} with $R$ derived from \eqref{eq:final} or \eqref{eq:final2}, depending on the number of dimensions. The Newtonian potential is that of baryonic matter made of gas, dust and stars. Solving numerically the referenced equations and comparing the results with observation will give the expression of $R$ as a function of $r$ and possibly $z$ and constrain the unknowns $L_2, n$ and the asymptotic value $R(\infty)=r_0$. Alternatively, one could use the expression for $R$ we derived in Section \ref{sec:rot} using a NFW profile for the effective potential, but in this case there will be no information on $n$.

\subsection{Motion of S stars}

It is known that there are several stars orbiting the Sgr A$^{*}$ black hole, see, for example, \cite{sstar,sstar2,sstar3,sstar4}. We propose to compare with observations the predicted orbit of the star S2, including the effect of the radius $R$. For other similar works see \cite{sstar5,sstar6} and references therein. The geodesics can be predicted using the modified Schwarzschild metric
\begin{equation}
    ds^2=-\left(1-\dfrac{2M}{r} \right)\, dt^2+\left(1-\dfrac{2M}{r} \right)^{-1}\, dr^2+r^2\,d\Omega_2+R(r)^2\,d\Omega_n
\end{equation}
where $d\Omega_i$ is the metric of a $i$-dimensional hypersphere. The geodesics on the equatorial plane ($\theta=\pi/2$) are\footnote{The Christoffel symbols where calculated with xAct \url{https://www.xact.es/}.}
\begin{align}
    \ddot{t}&=-\dfrac{M}{2Mr-r^2}\,\dot{t}\dot{r},\\
    \ddot{r}&=\dfrac{M(r-2M)}{r^3}\,\dot{t}^2+\dfrac{M}{2Mr-r^2}\,\dot{r}^2+(2M-r)\,\dot{\phi}^2+\dfrac{2M-r}{r}\,R\,R^\prime\,(\dot{\theta}_1^2+\dot{\theta}_2^2\,\sin^2\theta_1),\\
    \ddot{\phi}&=\dfrac{1}{r}\,\dot{r}\,\dot{\phi},\\
    \ddot{\theta}_1&=\dfrac{R^\prime}{R}\,\dot{r}\,\dot{\theta}_1-\dfrac{1}{2}\,\sin(2\theta_1)\,\dot{\theta}_2^2,\\
    \ddot{\theta}_2&=\dfrac{R^\prime}{R}\,\dot{r}\,\dot{\theta}_2-\cot\theta_1\,\dot{\theta}_1\dot{\theta}_2;
\end{align}
we notice that the extra dimensions modify the radial geodesic, while, in turn, the geodesics relative to extra dimensions depend on $r$; a dot indicate a derivative with respect to the affine parameter, a prime a derivative with respect to $r$. Since Sgr A$^{*}$ is, with good approximation, at the center of the Galaxy and therefore of the fake halo, we could use for $R$ the expression 
\begin{equation}\label{eq:expr}
    R(r)=L_2\,k\,r^{1/2}, \qquad k=\dfrac{1}{\sqrt{4\pi G\,\rho_0}}
\end{equation}
(compare to equation \eqref{eq:small}) where $r$ is the distance from the BH. The value of $k$ is known from studies of the dark matter halo of the Milky Way \cite{central}, while $L_2$ is not known and should be fitted. With this ansatz for $R$, the radial geodesic and those relative to the extra dimensions become
\begin{align}
     \ddot{r}&=\dfrac{M(r-2M)}{r^3}\,\dot{t}^2+\dfrac{M}{2Mr-r^2}\,\dot{r}^2+(2M-r)\,\dot{\phi}^2+\dfrac{2M-r}{r}\,\dfrac{L_2^2k^2}{2}\,(\dot{\theta}_1^2+\dot{\theta}_2^2\,\sin^2\theta_1),\\
     \ddot{\theta}_1&=\dfrac{1}{2r}\,\dot{r}\,\dot{\theta}_1-\dfrac{1}{2}\,\sin(2\theta_1)\,\dot{\theta}_2^2,\\
    \ddot{\theta}_2&=\dfrac{1}{2r}\,\dot{r}\,\dot{\theta}_2-\cot\theta_1\,\dot{\theta}_1\dot{\theta}_2.
\end{align}

We have derived the above expression in the special case of 6 dimensions, but they can easily be generalized to other dimension numbers: we notice that this experiment could also give an hint on the number of extra dimensions, simply by comparing geodesics calculated with different $n$.

\subsection{Black hole shadow}
\label{sec:shad}

The large gravitational field of a black hole can bend light and make it travel on circular orbits. Since the black hole does not emit light, in pictures such as those obtained by the Event Horizon Telescope \cite{EHT,EHT2,EHT3} it appears as a black disc, known as the black hole shadow. This shadow is observable \cite{shadow,shadow2} and could be used to test General Relativity and, in particular, our model.

The metric we consider in this Subsection is a modified Kerr metric
\begin{equation}
    ds^2=-\left( 1-\dfrac{2M r}{\Sigma} \right)dt^2-\dfrac{4Mr\,a\,\sin^2\theta}{\Sigma}\,dtd\phi+\dfrac{\Sigma}{\Delta}dr^2+\Sigma d\theta^2+\left(r^2+a^2+\dfrac{2Mra^2\sin^2\theta}{\Sigma}\right)\sin^2\theta+R^2\,d\Omega_d,
\end{equation}
where, as usual, 
\begin{equation*}
    \Delta=r^2+a^2-2Mr, \qquad \Sigma=r^2+a^2\cos^2\theta
\end{equation*}
The procedure to calculate the geodesics is outlined in the Appendix \ref{sec:geod}, here we only report their expression in the case of 6 dimensions:
\begin{align}
    \Sigma\,\dot{t}&=-a\,\left( a\,E\,\sin^2\theta-L_z \right)+\dfrac{r^2+a^2}{\Delta}\,(E\,(r^2+a^2)-L_z\,a),\\
    \Sigma^2\,\dot{r}^2&=\left( E(r^2+a^2)-L_z\,a \right)^2-\Delta\,\left( L^2+\dfrac{r^2\,L_2^2}{R^2} \right),\\
    \Sigma^2\,\dot{\theta}^2&=L^2-\cos^2\theta\,\left( \dfrac{L_z^2}{\sin^2\theta}-a^2e \right),\\
    \Sigma\,\dot{\phi}&=-\left( a\,E-\dfrac{L_z}{\sin^2\theta} \right)+a\,\left( E\,(r^2+a^2)-L_z\,a \right),\\
    \Sigma\,\dot{\theta}_1&=-\dfrac{r^2}{R^2}\,\sqrt{L_2^2-\dfrac{L_1^2}{\sin^2\theta_1}},\\
    \Sigma\,\dot{\theta}_2&=\dfrac{L_1}{\sin^2\theta_1}\,\dfrac{r^2}{R^2};
\end{align}
here $E$ is the energy of the particle, $m$, its mass, $L$ and $L_z$ are the total angular momentum and its third components, $L_1$ is a sort of angular momentum relative to the extra dimensions and $L_2$ is a separation constant. We notice again that the extra dimensions modify the radial geodesic, compare to \cite{BH}, Section 3.6.2. Following \cite{shadow4}, we consider the tetrad
\begin{align}
    e_0&=\dfrac{(r^2+a^2)\,\partial_t+a\,\partial_\phi}{\sqrt{\Sigma\,\Delta}},\\
    e_1&=\dfrac{\partial_\theta}{\sqrt{\Sigma}},\\
    e_2&=\dfrac{-\partial_\phi-a\,\sin^2\theta\,\partial_t}{\sqrt{\Sigma\, \Delta}},\\
    e_3&=-\sqrt{\dfrac{\Delta}{\Sigma}}\,\partial_r,\\
    e_4&=R\,\partial_{\theta_1},\\
    e_5&=R\,\sin\theta_1\,\partial_{\theta_2}.
\end{align}
We consider the vector at the emission site
\begin{equation}
    \lambda_E=\left.\dot{t}\,\partial_t+\dot{r}\,\partial_r+\dot{\theta}\,\partial_\theta+\dot{\phi}\,\partial_\phi+\dot{\theta}_1\,\partial_{\theta_1}+\dot{\theta}_2\,\partial_{\theta_2}\right|_{r=r_E},
\end{equation}
and the vector at the observation site
\begin{equation}
    \lambda_O=\left.\dfrac{a\,L_z-(a^2+r^2)}{\sqrt{\Sigma\,\Delta}}\,\left( -e_0+\cos\Psi\sin\Theta e_1+\sin\Psi\sin\Theta e_2+\cos\Theta e_3+e_4+e_5 \right)\right|_{r=r_O}
\end{equation}
where $\Theta$ and $\Phi$ are the celestial coordinates of the shadow and compare the coefficients of $\partial_r$ and $\partial_\phi$, thus obtaining the somewhat complicated expression
\begin{equation}
\begin{split}
    \cos\Theta&=\Bigg( \dfrac{\Delta_O\,\Sigma_O\,\Sigma_E}{\Delta_E}\,\left( 2\,(2a^2+r_O^2+r_E^2 \right)^2+L_{2E}^2\,\Delta_E\,(R(r_E)-r_E\,R^\prime(r_E))^2-\dfrac{4r_O^2\,L_{2e}^2\,\Delta_O\,R(r_E)^6}{R(r_O)^2}-L_{2e}^2\,\Delta_O\,r_E^2\,R(r_E)^4+\\
    &+L_{2e}^4\,\Delta_O\,\Delta_E\,(R(r_E)-r_E\,R^\prime(r_E))\Bigg)^{1/2}\,\left( 2(2a^2\,r_E^2+r_O^2)\,R(r_E)^3-L_{2e}^2\,\Delta_E\,(R(r_E)-r_E\,R(^\prime(r_E)) \right)^{-1}
\end{split}
\end{equation}
\begin{equation}
    \begin{split}
        \sin\Psi&=-\Bigg( a^2\,\Delta_E\,\Sigma_O\,\Bigg( \Delta_E\Sigma_O-\Delta_O\Sigma_E+ \Big(L_{2e}^2\,\Delta_O^2\,\Sigma_E\,\Big( -4\,(r_E^2\,R(r_O)^2-r_O^2\,R(r_E)^2)\,R(r_E)^4+\\
        &+L_{2e}^2\,\Delta_E\,R(r_o)^2\,\left( R(r_E)-r_E\,R^\prime(r_E) \right) \Big)\Big)\,\Big(R(r_O)^2\,\Big( 2(2a^2+r_O^2+r_E^2)\,R(r_E)^3+\\
        &+L_{2e}^2\, \Delta_E\,\Big( R(r_E)-r_E\,R^\prime(r_E) \Big)\Big)\Big)^{-1} \Bigg)^{-1} \Bigg)^{1/2}+\\
        &+\Delta_E\,\sqrt{\Delta_O\,\Sigma_E}\,\Bigg( -2a^2\,(2a^2+r_E^2+r_O^2-\Delta_O)\,R(r_E)^3+a^2\,L_{2e}^2\Delta_E\,(R(r_E)-r_E\,R^\prime(r_E))+\\
        &+\dfrac{1}{\sin^2\theta_O}\Big( 2(a^2+r_E)^2\,\Delta_O\,R(r_E)^3-L_{2e}^2\,\Delta_O\,\Delta_E\,(R(r_E)-r_E\,R^\prime(r_E))\ \Big) \Bigg)\times\\
        &\times\Bigg( a\,\Delta_O\,\Big( -2(2a^2+r_O^2+r_E^2)\,R(r_E)^3+L_{2e}^2\,\Delta_E\,(R(r_E)-r_E\,R^\prime(r_E)) \Big) \, \Big( \Delta_E\Sigma_O-\Delta_O\Sigma_E+\\
        &+\,L_{2e}^2\,\Delta_O^2\,\Sigma_E\,\Big( -4\,(r_E^2\,R(r_O)^2-r_E\,R(r_O)^2)\,R(r_E)^4+L_{2e}^2\,\Delta_E\,(R(r_E)-r_E\,R^\prime(r_E))^2\,R(r_O) \Big)\,R(r_O)^{-2}\\
        &\Big( 2\,(2a^2+r_E^2+r_O^2)\,R(r_E)^3-\Delta_E\,L_{2e}^2 \,(R(r_E)-r_E\,R^\prime(r_E))\Big)^{-1} \Big)^{1/2}  \Bigg)^{-1}
    \end{split}
\end{equation}
above, we have $\Delta_E=r_E^2-2Mr_E+a^2$, $\Sigma_E=r_E^2+a^2\cos^2\theta_E$ and analogously for $\Delta_O$ and $\Sigma_O$.

If we consider the expression \eqref{eq:expr} for $R$, we obtain the simpler expressions
\begin{equation}
    \begin{split}
        \cos\Theta&=\sqrt{\dfrac{\Delta_O\,\Sigma_E}{\Delta_E\,\Sigma_O}}\,\Bigg( 16\,k^4\,r_E^2\,(2a^2+r_E^2+r_O^2)^2\,\Delta_E\,(\Delta_p-\Delta_O)+8\,k^2\,r_E\,\Big(2\,r_E(r_E-r_O)\,\Delta_O-(2a^2r_E^2+r_O^2)\,\Delta_E\Big) \Bigg)^{1/2}\\
        &\Big( 4k^2\,r_E\,(2a^2+r_E^2+r_O^2)-\Delta_E \Big)^{-1}
    \end{split}
\end{equation}
\begin{equation}
    \begin{split}
        \sin\Psi&=-\Bigg( a^2\,\Delta_E\Sigma_O\,\Big( \Delta_E\Sigma_O-\Delta_O\Sigma_E+\dfrac{\Delta_O^2\,\Sigma_E\,(16\,k^2\,r_E^2\,(r_E-r_O)+\Delta_E)}{\Delta_E-4k^2\,r_E\,(2a^2+r_E^2+r_O^2)} \Big)^{-1} \Bigg)^{1/2}+\\
        &+\sqrt{\dfrac{\Sigma_E}{a^2\,\Delta_O}}\,\Bigg( a^2\,\Delta_E\,\Big( \Delta_E-4k^2\,r_E\,(2a^2+r_E^2+r_O^2-\Delta_O) \Big) + \dfrac{\Delta_O\,\Delta_E}{\sin^2\theta_O}\,\Big( 4k^2\,r_E\,(a^2+r_E^2)-\Delta_E \Big) \Bigg)\times\\
        &\times\Bigg( \Delta_E-4k^2\,r_E\,(2a^2+r_E^2+r_O^2)\,\Bigg( \Delta_E\Sigma_O-\Sigma_E\Delta_O+\dfrac{\Delta_O^2\,\Sigma_E\,(16\,k^2\,r_E^2\,(r_E-r_O)+\Delta_E)}{\Delta_E-4k^2\,r_E\,(2a^2+r_E^2+r_O^2)} \Bigg)^{1/2} \Bigg)^{-1}.
    \end{split}
\end{equation}
We can further assume that $k$ is small, or a deviation from usual theory would have been observed, so we expand to the first order in $k$
\begin{equation}
    \cos\Theta=-\dfrac{1}{\Delta_E}\,\sqrt{\dfrac{(\Delta_E-\Delta_O)\,\Delta_O\,\Sigma_E}{\Sigma_O}}+4\,k^2\,\Bigg[ \dfrac{r_E\,\Delta_O^2\,\Sigma_E\,(2a^2+r_O^2+2r_Or_E-r_E^2)}{\Delta_E^2\,\sqrt{\Delta_O\,\Sigma_O\,\Sigma_E\,(\Delta_E-\Delta_O)}}  \Bigg] +O(k^4)
\end{equation}
\begin{equation}
    \begin{split}
        \sin\Psi&=-\dfrac{a\,\Delta_E\,\Sigma_O}{\sqrt{\Sigma_O\,(\Delta_E^2\Sigma_O^2+\Delta_O^2\Sigma_E-\Delta_O\Delta_E\Sigma_E)}}+\dfrac{\Delta_E\,\sqrt{\Delta_O\Delta_E\Sigma_E}(a^2-\Delta_O\sin^{-2}\theta_O)}{a\,\Delta_O\,{\sqrt{\Sigma_O\,(\Delta_E^2\Sigma_O+\Delta_O^2\Sigma_E-\Delta_O\Delta_E\Sigma_E)}}}+\\
        &+4r_E\,k^2\,\Bigg[ \dfrac{a\,(2a^2+r_O^2+2r_Or_E-r_E^2)\,\sqrt{\Sigma_O}}{(\Delta_E^2\Sigma_O+\Delta_O^2\Sigma_E-\Delta_O\Delta_E\Sigma_E)^{3/2}}+\dfrac{\sqrt{\Delta_O\Delta_E\Sigma_E}}{a\,(\Delta_E^2\Sigma_O+\Delta_O^2\Sigma_E-\Delta_O\Delta_E\Sigma_E)^{3/2}} \times\\
        &\times\Big( a^2\,(\Delta_E^2\Sigma_O-\Delta_O\,(2a^2+r_O^2+2r_Or_E-r_E^2-\Delta_o+\Delta_E)\,\Sigma_E +\\
        &+\sin^{-2}\theta_O\,\Big(-(a^2+r_O^2)\,\Delta_E^2\Sigma_O+\Delta_O\,\Sigma_E\,\Big(\Delta_O\,(a^2+2r_Or_E-r_E^2)+\Delta_E\,(a^2+r_O^2)\Big)\Big)\Bigg]+O(k^4)
    \end{split}
\end{equation}
Following the procedure explained in \cite{shadow4}, one can derive the shape of the shadow and compare it with observational data.

\subsection{Time delays}

In order to study the time delay, we consider the metric \eqref{eq:met} and the geodesics
\begin{align}
    \dot{r}^2&=(1-2\Phi)\,E^2-\dfrac{L_z^2}{r^2}-\dfrac{L_2^2}{R(r)^2},\\
    \dot{t}&=E\,(1-2\Phi).
\end{align}
We calculate the ratio
\begin{equation}
    \dfrac{\dot{t}}{\dot{r}}=\dfrac{dt}{dr}=\dfrac{E\,(1-2\Phi)}{\sqrt{(1-2\Phi)\,E^2-\dfrac{L_z^2}{r^2}-\dfrac{L_2^2}{R(r)^2}}}.
\end{equation}
We integrate the above expression and obtain
\begin{equation}
    \Delta t=\int \dfrac{(1-2\Phi)\,dr}{\sqrt{(1-2\Phi)-\dfrac{b^2}{r^2}-\dfrac{\ell_2^2}{R(r)^2}}},
\end{equation}
where $b=L_z/E$ and $\ell_2=L_2/E$ are impact parameters.

We assume that the term $\ell_2^2/R^2$ is small, or we would have already noticed its effect, thus we expand at the first order in this term:
\begin{subequations}
    \begin{equation}
        \Delta t=\Delta t_0+\Delta t_R +O(R)^{-2}
    \end{equation}
    \begin{equation}
        \Delta t_0=\int \dfrac{(1-2\Phi)\,dr}{\sqrt{(1-2\Phi)-\dfrac{b^2}{r^2}}},
    \end{equation}
    \begin{equation}
        \Delta t_R=-\int  \dfrac{dr \,(1-2\Phi)}{\left((1-2\Phi)-\dfrac{b^2}{r^2}\right)^{3/2}}\, \dfrac{\ell_2}{2R(r)}.
    \end{equation}
\end{subequations}
In place of $R$ one should substitute the proper analytic value, equation \eqref{eq:def_R}, taking into consideration the distance of the source (and of the observer) from the center of the fake halo, and compare the result with the measured value of the delay. We notice that with expression \eqref{eq:def_R}, one could also obtain an estimate of the asymptotic value $r_0$.

\section{Conclusion and discussion}
\label{sec:concl}
In this paper, we have presented a simple model of a higher dimensional spacetime with extra dimension compactified to a hypersphere and assumed that its radius $R$ depends on the radial coordinate $r$. We have found that the radius is affected by the presence of mass in the 4d submanifold and by the Newtonian potential it generates and we were able to show that $R^{-2}$ behaves as an additional potential that affects the motion of matter in the macroscopic 4d spacetime. We have associated this potential with the phenomenon of dark matter, proving that we can explain the rotation curve of spiral galaxies and give an explicit form of $R$. With our model, we can explain the presence of galaxies without dark matter, but we need further studies to explain dark galaxies and the Bullet Cluster; in particular, we need to study the Newtonian potential of these objects and the distribution of matter through gravitational lensing. We have also presented six possible experiments to verify our hypothesis: the study of the velocity of satellite galaxies; the study of elliptical and spiral galaxies; the motion of the S stars around Sgr A$^{*}$; the shadow of Sgr A$^{*}$, or, in principle, of any supermassive black hole, in particular of M87$^{*}$; and finally, the study of time delays. It is possible to impose constraints on  the number of extra dimensions $n$ by comparing the theoretical results with observations in the first four experiments. These experiments are not new, but they can be performed with today technology.

\begin{figure}
    \centering
    \subfigure{\includegraphics[width=0.6\linewidth]{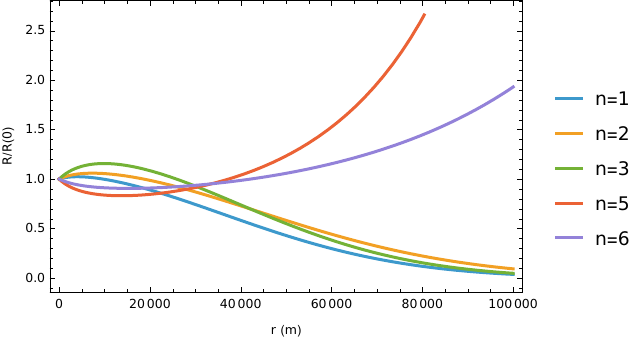}}
    \subfigure{\includegraphics[width=0.6\linewidth]{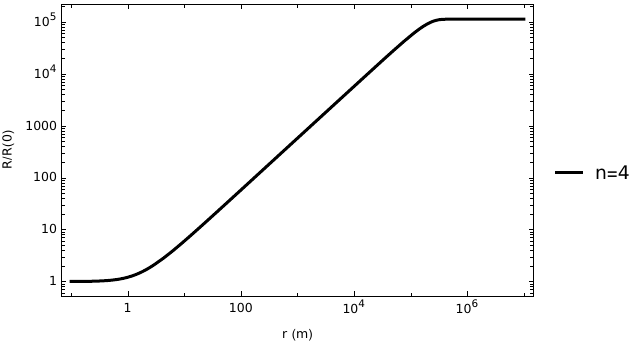}}
    \caption{The evolution of the radius $R$ around a neutron star of $1.5$ M$_\Sun$ as a function of the distance for different number of extra dimensions. Top panel: case of $n=\{1,2,3,5,6\}$; Bottom panel: case $n=4$.}
    \label{fig:NS}
\end{figure}

As a final note, we point out that, in principle, equations \eqref{eq:final} and \eqref{eq:final2} predict that the value of $R$ around any concentration of mass will be different with respect to the value in vacuum: we have checked that the potential of the Earth, the Moon and the Sun is too small to give values of $R$ much different from the asymptotic one; this also implies that any ground experiment which tries to measure the size and the number of the extra dimensions needs to probe distances of the order of fm, or smaller \cite{ap1,ap2,ap3,ap4}: to the best of our knowledge no such experiment has been proposed yet. In the case of a neutron star, we report in Figure \ref{fig:NS} the evolution of $R/R(0)$, where $R(0)$ is the value of $R$ on the surface of the star for $n=\{1,2,3,5,6\}$ in the top panel and $n=4$ in the bottom panel. We notice the complicated behavior of the radius $R$: for $n\leq3$ the radius tends to smaller and smaller values as the distance from the surface increases; for $n\geq5$, the situation is the opposite: $R$ will grow with $r$. However, closer to the surface the situation changes: for $n\leq3$ $R$ increases by about 20\%, while for $n\geq5$ it decreases by the same amount. In the case $n=4$, we notice from the bottom panel of the Figure, that $R$ will reach a large asymptotic value, while for $n\geq5$, we have found that $R$ grows indefinitely.  We leave this fact here in case someone more clever than the present author can come up with an experiment to measure $R$ around a neutron star. The perturbation method we have used is not valid around objects with larger potentials such as black holes; however, one might wonder whether the extra dimensions will be significantly modified around such large gravitational fields and what happens close to singularities.

The following paper of the series \cite{mio2} will deal with the collapse of large scale structures, in which dark matter gives an important contribution.

Together with the companion paper \cite{mio2} and with papers in the literature such as \cite{RS1,RS2,RS3,RS4,RS5,RS6}, we can give a single explanation to the dark sector of the Universe: the presence of extra dimensions. We try to substitute something rather obscure (dark matter and dark energy) with something for which we do not have experimental evidence; however, we believe that this model, because of its economy, needs further studies.

\appendix
\section{Einstein equations}
\label{sec:ein}
In this Appendix, we derive equation \eqref{eq:final}.

Using xAct, we can find that the Einstein equations of the metric \eqref{eq:met} for generic number of extra dimension $n$ are
\begin{subequations}
    \begin{equation}\label{eq:prima}
        -n\,\dfrac{R^{\prime\prime}}{R}-\dfrac{2n}{r}\,\dfrac{R^\prime}{R}-\dfrac{n(n-1)}{2}\,\left[ -\dfrac{1}{R^2}+\left( \dfrac{R^\prime}{R} \right)^2 \right]=8\pi G\,\rho\,(1-2\Phi),
    \end{equation}
    \begin{equation}\label{eq:seconda}
       n\,\Phi^\prime\,\left(\dfrac{R^\prime}{R}+\dfrac{2}{r}\right)+\dfrac{2\,\Phi^\prime}{r}+\dfrac{n(n-1)}{2}\,\left[ -\dfrac{1}{R^2}+\left( \dfrac{R^\prime}{R} \right)^2 \right]=0,
    \end{equation}
    \begin{equation}\label{eq:terza}
        n\,\dfrac{R^{\prime\prime}}{R}+\Phi^{\prime\prime}+\dfrac{\Phi^\prime}{r}+n\,\dfrac{R^\prime}{R}\,\left( \dfrac{1}{r}+\Phi^\prime \right)+\dfrac{n(n-1)}{2}\,\left[ -\dfrac{1}{R^2}+\left( \dfrac{R^\prime}{R} \right)^2 \right]=0,
    \end{equation}
    \begin{equation}\label{eq:quarta}
        \Phi^{\prime\prime}+\dfrac{2\,\Phi^\prime}{r}+(n-1)\,\dfrac{R^{\prime\prime}}{R}+(n-1)\,\dfrac{R^\prime}{R}\,\left( \dfrac{2}{r}+\Phi^\prime \right)+\dfrac{n(n-1)}{2}\,\left[ -\dfrac{1}{R^2}+\left( \dfrac{R^\prime}{R} \right)^2 \right]=0,
    \end{equation}
\end{subequations}
where we have used dust as a source and considered only terms up to $O(\Phi)$.

We consider the linear combination
\begin{equation}\label{eq:comb}
    a\,(A1a)+b\,(A1b)+c\,(A1c)+d\,(A1d).
\end{equation}
We impose
\begin{equation}
    a=\dfrac{r}{4n-n^2},\qquad b=\dfrac{r}{n(n-4)}, \qquad c=\dfrac{(n-2)\,r}{(n-4)\,n}, \qquad d=\dfrac{r}{4-n},
\end{equation}
and obtain
\begin{equation}
    \dfrac{R^\prime}{R}=\dfrac{8\pi\,G\,\rho\,(1-2\Phi)-n\,\Phi^\prime-2r\,\Phi^{\prime\prime}}{n(n-4)}
\end{equation}
which is equation \eqref{eq:final}. In order to derive the case for $n=4$, we impose this value in equation \eqref{eq:comb} and impose
\begin{equation}
    a=-\dfrac{1}{6},\qquad b=\dfrac{1}{6}, \qquad c=-\dfrac{2}{3}, \qquad d=\dfrac{2}{3},
\end{equation}
and obtain
\begin{equation}
    \left( \dfrac{R^\prime}{R} \right)^2+2\,\dfrac{R^\prime}{R}-\dfrac{1}{R^2}=-\dfrac{8\pi G\,\rho}{12}\,\left( 1-2\Phi \right)-\dfrac{\Phi^\prime}{2r}
\end{equation}
which is equation \eqref{eq:final2}. There are two branches in the above equation: one should choose the one that gives positive real numbers. We have chosen the above parameters in order to eliminate the second derivative $R^{\prime\prime}$; in this way while numerically solving the equations, we only need one boundary condition.

\section{Derivation of the Kerr geodesics}
\label{sec:geod}
In this appendix, we derive the geodesics of the modified Kerr metric reported in Section \ref{sec:shad}. We use a Hamilton-Jacobi approach. We start from the Hamilton-Jacobi equation
\begin{equation}
    -\dfrac{dS}{d\Lambda}=\dfrac{1}{2}\,g^{\mu\nu}\,\dfrac{dS}{dx^\mu}\,\dfrac{dS}{dx^\nu}
\end{equation}
which, in the case of 6 dimensions, is rewritten as
\begin{equation}
\begin{split}
    -\dfrac{dS}{d\lambda}&=-\dfrac{1}{2\Sigma\Delta}\,\left( (r^2+a^2)\,\dfrac{dS}{dt}+a\,\dfrac{dS}{d\phi} \right)^2+\dfrac{1}{2\Sigma\,\sin^2\theta}\,\left( \dfrac{dS}{d\phi} -a\,\sin^2\theta\,\dfrac{dS}{dt} \right)^2+\dfrac{\Delta}{2\Sigma}\,\left( \dfrac{dS}{dr} \right)^2+\dfrac{1}{2\Sigma}\,\left( \dfrac{dS}{d\theta} \right)^2+\\
    &+\dfrac{1}{2R^2}\,\left( \left(\dfrac{dS}{d\theta_1} \right)^2 + \dfrac{1}{\sin^2\theta_1}\,\left( \dfrac{dS}{d\theta_2} \right)^2 \right).
\end{split}
\end{equation}
We make the ansatz 
\begin{equation}
    S=\dfrac{m^2}{2}\,\lambda-E\,t+L_z\,\phi+L_1\,\theta_2+S_r(r)+S_{\theta}(\theta)+S_{\theta_1}(\theta_1).
\end{equation}
Substituting this expression in the above equation, we find
\begin{equation}
    \begin{split}
        \dfrac{(-a^2E+aL_z+r^2)\,R^2}{\Delta}&=\Delta\,R^2\,\left( \dfrac{dS_r}{dr} \right)^2+ R^2\,\left[ -2aEL_z+m^2r^2+a^2m^2\,\cos^2\theta+a^2E^2\sin^2\theta+\left( \dfrac{dS_\theta}{d\theta} \right)^2 \right]+\\
        &+\Sigma\,\left( \dfrac{L_1^2}{\sin^2\theta_1}+\left( \dfrac{dS_{\theta_1}}{d\theta_1} \right)^2 \right).
    \end{split}
\end{equation}
This is separable
\begin{subequations}\label{eq:action}
    \begin{equation}
        \dfrac{dS_r}{dr}=\dfrac{1}{\Delta}\,\sqrt{(-a^2E+aL_z+r^2)^2-\Delta\,\left( m^2r^2+L^2+\dfrac{L_2^2r^2}{R^2} \right)}=\dfrac{\sqrt{\mathcal{R}}}{\Delta},
    \end{equation}
    \begin{equation}
        \dfrac{dS_\theta}{d\theta}=\sqrt{L^2-m^2a^2\cos^2\theta-\left( Ea\,\sin\theta-\dfrac{L_z}{\sin\theta} \right)^2}=\sqrt{\Theta},
    \end{equation}
    \begin{equation}
        \dfrac{dS_{\theta_1}}{d\theta_1}=\sqrt{L_2^2-\dfrac{L_1^2}{\sin^2\theta_1}}=\sqrt{\Theta_1}.
    \end{equation}
\end{subequations}
Now, using the above expressions, we derive the action in $L^2$, $L_z$, $L_1$, $L_2$, $E$, and $m^2$, obtaining, respectively
\begin{subequations}
    \begin{equation}\label{eq:zero}
        \dfrac{dS}{dL}=0=\int \dfrac{dr}{\sqrt{\mathcal{R}}}-\int \dfrac{d\theta}{\sqrt{\Theta}},
    \end{equation}
    \begin{equation}
        \dfrac{dS}{dL_z}=\phi=\int dr\,\left( \dfrac{d}{dL_z}\,\dfrac{\sqrt{\mathcal{R}}}{\Delta} \right)+\int d\theta\,\left( \dfrac{d\sqrt{\Theta}}{dL_z} \right)
    \end{equation}
    \begin{equation}
        \dfrac{dS}{dL_1}=\theta_2=\int d\theta_1\left( \dfrac{d\sqrt{\Theta_1}}{dL_1} \right)
    \end{equation}
    \begin{equation}
        \dfrac{dS}{dL_2}=0=\int dr\,\left( \dfrac{d}{dL_2}\,\dfrac{\sqrt{\mathcal{R}}}{\Delta} \right)+\int d\theta\,\left( \dfrac{d\sqrt\Theta_1}{dL_2} \right)
    \end{equation}
    \begin{equation}
        \dfrac{dS}{dE}=t=\int dr\,\left( \dfrac{d}{dE}\,\dfrac{\sqrt{\mathcal{R}}}{\Delta} \right)+\int d\theta\,\left( \dfrac{d\sqrt{\Theta}}{dE} \right)
    \end{equation}
    \begin{equation}\label{eq:lam}
        \dfrac{dS}{dm^2}=\lambda=\int dr\,\left( \dfrac{d}{dm^2}\,\dfrac{\sqrt{\mathcal{R}}}{\Delta} \right)+\int d\theta\,\left( \dfrac{d\sqrt{\Theta}}{dm^2} \right)
    \end{equation}
\end{subequations}
Combining the above equations, we obtain the geodesics reported in Section \ref{sec:shad}; for example, we first differentiate \eqref{eq:lam} and obtain
\begin{equation}
    1=\dfrac{dr}{d\lambda}\,\dfrac{dS_r}{dm^2}+\dfrac{d\theta}{d\lambda}\,\dfrac{dS_\theta}{dm^2},
\end{equation}
we further derive equation \eqref{eq:zero} in $\lambda$ and using the expression we have just found and equations \eqref{eq:action} we can derive the expressions for $\dot{r}$ and $\dot{\theta}$. 

\bibliography{biblio}

@ARTICLE{ap1,
       author = {{Horv{\'a}th}, Anna and {Forg{\'a}cs-Dajka}, Emese and {Barnaf{\"o}ldi}, Gergely G{\'a}bor},
        title = "{Application of Kaluza-Klein theory in modelling compact stars: exploring extra dimensions}",
      journal = {Monthly Notices of Royal Astronomical of Science},
         year = 2025,
        month = jan,
       volume = {536},
       number = {1},
        pages = {816-826},
          doi = {10.1093/mnras/stae2637},
archivePrefix = {arXiv},
       eprint = {2408.16497},
 primaryClass = {gr-qc},
       adsurl = {https://ui.adsabs.harvard.edu/abs/2025MNRAS.536..816H}
}

@ARTICLE{ap2,
       author = {{Clarkson}, Chris and {Maartens}, Roy},
        title = "{Gravity-wave detectors as probes of extra dimensions}",
      journal = {General Relativity and Gravitation},
         year = 2005,
        month = oct,
       volume = {37},
       number = {10},
        pages = {1681-1687},
          doi = {10.1007/s10714-005-0150-8},
archivePrefix = {arXiv},
       eprint = {astro-ph/0505277},
 primaryClass = {astro-ph},
       adsurl = {https://ui.adsabs.harvard.edu/abs/2005GReGr..37.1681C}
}

@ARTICLE{ap3,
       author = {{Clarkson}, Chris and {Maartens}, Roy},
        title = "{Gravity-Wave Detectors as Probes of Extra Dimensions}",
      journal = {International Journal of Modern Physics D},
         year = 2005,
        month = jan,
       volume = {14},
       number = {12},
        pages = {2347-2353},
          doi = {10.1142/S0218271805007905},
       adsurl = {https://ui.adsabs.harvard.edu/abs/2005IJMPD..14.2347C}
}

@article{ap4,
  title = {Limiting the number of extra dimensions with shortcuts},
  author = {Lin, Zi-Chao and Yu, Hao and Gong, Yungui},
  journal = {Phys. Rev. D},
  volume = {109},
  issue = {10},
  pages = {104015},
  numpages = {9},
  year = {2024},
  month = {May},
  publisher = {American Physical Society},
  doi = {10.1103/PhysRevD.109.104015},
  url = {https://link.aps.org/doi/10.1103/PhysRevD.109.104015}
}

@unpublished{mio2,
    author = {Mattia Villani},
    title = {Cosmological extra dimensions can mimic dark matter II. Growth of cosmological
structures},
    note = {Submitted to Annals of Physics}
}

@ARTICLE{NoDM,
       author = {{Guo}, Qi and {Hu}, Huijie and {Zheng}, Zheng and {Liao}, Shihong and {Du}, Wei and {Mao}, Shude and {Jiang}, Linhua and {Wang}, Jing and {Peng}, Yingjie and {Gao}, Liang and {Wang}, Jie and {Wu}, Hong},
        title = "{Further evidence for a population of dark-matter-deficient dwarf galaxies}",
      journal = {Nature Astronomy},
         year = 2020,
        month = jan,
       volume = {4},
        pages = {246-251},
          doi = {10.1038/s41550-019-0930-9},
archivePrefix = {arXiv},
       eprint = {1908.00046},
 primaryClass = {astro-ph.GA},
       adsurl = {https://ui.adsabs.harvard.edu/abs/2020NatAs...4..246G}
}

@ARTICLE{DF2,
       author = {{van Dokkum}, Pieter and {Danieli}, Shany and {Cohen}, Yotam and {Merritt}, Allison and {Romanowsky}, Aaron J. and {Abraham}, Roberto and {Brodie}, Jean and {Conroy}, Charlie and {Lokhorst}, Deborah and {Mowla}, Lamiya and {O'Sullivan}, Ewan and {Zhang}, Jielai},
        title = "{A galaxy lacking dark matter}",
      journal = {Nature},
         year = 2018,
        month = mar,
       volume = {555},
       number = {7698},
        pages = {629-632},
          doi = {10.1038/nature25767},
archivePrefix = {arXiv},
       eprint = {1803.10237},
 primaryClass = {astro-ph.GA},
       adsurl = {https://ui.adsabs.harvard.edu/abs/2018Natur.555..629V}
}

@ARTICLE{DF2b,
       author = {{Guo}, Qi and {Hu}, Huijie and {Zheng}, Zheng and {Liao}, Shihong and {Du}, Wei and {Mao}, Shude and {Jiang}, Linhua and {Wang}, Jing and {Peng}, Yingjie and {Gao}, Liang and {Wang}, Jie and {Wu}, Hong},
        title = "{Further evidence for a population of dark-matter-deficient dwarf galaxies}",
      journal = {Nature Astronomy},
         year = 2020,
        month = jan,
       volume = {4},
        pages = {246-251},
          doi = {10.1038/s41550-019-0930-9},
archivePrefix = {arXiv},
       eprint = {1908.00046},
 primaryClass = {astro-ph.GA},
       adsurl = {https://ui.adsabs.harvard.edu/abs/2020NatAs...4..246G}
}

@ARTICLE{DF4,
       author = {{van Dokkum}, Pieter and {Danieli}, Shany and {Abraham}, Roberto and {Conroy}, Charlie and {Romanowsky}, Aaron J.},
        title = "{A Second Galaxy Missing Dark Matter in the NGC 1052 Group}",
      journal = {Ap. J. Letters},
         year = 2019,
        month = mar,
       volume = {874},
       number = {1},
          eid = {L5},
        pages = {L5},
          doi = {10.3847/2041-8213/ab0d92},
archivePrefix = {arXiv},
       eprint = {1901.05973},
 primaryClass = {astro-ph.GA},
       adsurl = {https://ui.adsabs.harvard.edu/abs/2019ApJ...874L...5V}
}

@ARTICLE{DF9,
       author = {{Keim}, Michael A. and {van Dokkum}, Pieter and {Shen}, Zili and {Danieli}, Shany and {Pasha}, Imad},
        title = "{A Third Galaxy Missing Dark Matter along a Trail of Galaxies in the NGC 1052 Field}",
      journal = {arXiv e-prints},
         year = 2026,
        month = mar,
          eid = {arXiv:2603.15860},
        pages = {arXiv:2603.15860},
          doi = {10.48550/arXiv.2603.15860},
archivePrefix = {arXiv},
       eprint = {2603.15860},
 primaryClass = {astro-ph.GA},
       adsurl = {https://ui.adsabs.harvard.edu/abs/2026arXiv260315860K}
}

@ARTICLE{FC224,
doi = {10.3847/2515-5172/ad7112},
url = {https://doi.org/10.3847/2515-5172/ad7112},
year = {2024},
month = {aug},
publisher = {The American Astronomical Society},
volume = {8},
number = {8},
pages = {202},
author = {Romanowsky, Aaron J. and Cabrera, Enrique and Janssens, Steven R.},
title = {A Candidate Dark Matter Deficient Dwarf Galaxy in the Fornax Cluster Identified through Overluminous Star Clusters},
journal = {Research Notes of the AAS}
}

@ARTICLE{MW,
       author = {{McMillan}, Paul J.},
        title = "{Mass models of the Milky Way}",
      journal = {MNRAS},
         year = 2011,
        month = jul,
       volume = {414},
       number = {3},
        pages = {2446-2457},
          doi = {10.1111/j.1365-2966.2011.18564.x},
archivePrefix = {arXiv},
       eprint = {1102.4340},
 primaryClass = {astro-ph.GA},
       adsurl = {https://ui.adsabs.harvard.edu/abs/2011MNRAS.414.2446M}
}

@INPROCEEDINGS{LMC,
       author = {{van der Marel}, Roeland P.},
        title = "{The Large Magellanic Cloud: structure and kinematics}",
    booktitle = {The Local Group as an Astrophysical Laboratory},
         year = 2006,
       editor = {{Livio}, Mario and {Brown}, Thomas M.},
       volume = {17},
        month = jan,
        pages = {47-71},
          doi = {10.48550/arXiv.astro-ph/0404192},
archivePrefix = {arXiv},
       eprint = {astro-ph/0404192},
 primaryClass = {astro-ph},
       adsurl = {https://ui.adsabs.harvard.edu/abs/2006lgal.symp...47V}
}

@ARTICLE{dark,
       author = {{Barrufet}, L. and {Oesch}, P.~A. and {Weibel}, A. and {Brammer}, G. and {Bezanson}, R. and {Bouwens}, R. and {Fudamoto}, Y. and {Gonzalez}, V. and {Gottumukkala}, R. and {Illingworth}, G. and {Heintz}, K.~E. and {Holden}, B. and {Labbe}, I. and {Magee}, D. and {Naidu}, R.~P. and {Nelson}, E. and {Stefanon}, M. and {Smit}, R. and {van Dokkum}, P. and {Weaver}, J.~R. and {Williams}, C.~C.},
        title = "{Unveiling the nature of infrared bright, optically dark galaxies with early JWST data}",
      journal = {MNRAS},
         year = 2023,
        month = jun,
       volume = {522},
       number = {1},
        pages = {449-456},
          doi = {10.1093/mnras/stad947},
archivePrefix = {arXiv},
       eprint = {2207.14733},
 primaryClass = {astro-ph.GA},
       adsurl = {https://ui.adsabs.harvard.edu/abs/2023MNRAS.522..449B}
}

@ARTICLE{dark2,
       author = {{P{\'e}rez-Gonz{\'a}lez}, Pablo G. and {Barro}, Guillermo and {Annunziatella}, Marianna and {Costantin}, Luca and {Garc{\'\i}a-Argum{\'a}nez}, {\'A}ngela and {McGrath}, Elizabeth J. and {M{\'e}rida}, Rosa M. and {Zavala}, Jorge A. and {Arrabal Haro}, Pablo and {Bagley}, Micaela B. and {Backhaus}, Bren E. and {Behroozi}, Peter and {Bell}, Eric F. and {Bisigello}, Laura and {Buat}, V{\'e}ronique and {Calabr{\`o}}, Antonello and {Casey}, Caitlin M. and {Cleri}, Nikko J. and {Coogan}, Rosemary T. and {Cooper}, M.~C. and {Cooray}, Asantha R. and {Dekel}, Avishai and {Dickinson}, Mark and {Elbaz}, David and {Ferguson}, Henry C. and {Finkelstein}, Steven L. and {Fontana}, Adriano and {Franco}, Maximilien and {Gardner}, Jonathan P. and {Giavalisco}, Mauro and {G{\'o}mez-Guijarro}, Carlos and {Grazian}, Andrea and {Grogin}, Norman A. and {Guo}, Yuchen and {Huertas-Company}, Marc and {Jogee}, Shardha and {Kartaltepe}, Jeyhan S. and {Kewley}, Lisa J. and {Kirkpatrick}, Allison and {Kocevski}, Dale D. and {Koekemoer}, Anton M. and {Long}, Arianna S. and {Lotz}, Jennifer M. and {Lucas}, Ray A. and {Papovich}, Casey and {Pirzkal}, Nor and {Ravindranath}, Swara and {Somerville}, Rachel S. and {Tacchella}, Sandro and {Trump}, Jonathan R. and {Wang}, Weichen and {Wilkins}, Stephen M. and {Wuyts}, Stijn and {Yang}, Guang and {Yung}, L.~Y. Aaron},
        title = "{CEERS Key Paper. IV. A Triality in the Nature of HST-dark Galaxies}",
      journal = {Ap. J. Letters},
         year = 2023,
        month = mar,
       volume = {946},
       number = {1},
          eid = {L16},
        pages = {L16},
          doi = {10.3847/2041-8213/acb3a5},
archivePrefix = {arXiv},
       eprint = {2211.00045},
 primaryClass = {astro-ph.GA},
       adsurl = {https://ui.adsabs.harvard.edu/abs/2023ApJ...946L..16P}
}

@article{dark3,
doi = {10.3847/2041-8213/adddab},
url = {https://doi.org/10.3847/2041-8213/adddab},
year = {2025},
month = {jun},
publisher = {The American Astronomical Society},
volume = {986},
number = {2},
pages = {L18},
author = {Li, Dayi (David) and Liu, Qing and Eadie, Gwendolyn M. and Abraham, Roberto G. and Marleau, Francine R. and Harris, William E. and van Dokkum, Pieter and Romanowsky, Aaron J. and Danieli, Shany and Brown, Patrick E. and Stringer, Alex},
title = {Candidate Dark Galaxy-2: Validation and Analysis of an Almost Dark Galaxy in the Perseus Cluster},
journal = {The Astrophysical Journal Letters}
}

@article{dark4,
doi = {10.1088/2041-8205/798/2/L45},
url = {https://doi.org/10.1088/2041-8205/798/2/L45},
year = {2015},
month = {jan},
publisher = {The American Astronomical Society},
volume = {798},
number = {2},
pages = {L45},
author = {van Dokkum, Pieter G. and Abraham, Roberto and Merritt, Allison and Zhang, Jielai and Geha, Marla and Conroy, Charlie},
title = {FORTY-SEVEN MILKY WAY-SIZED, EXTREMELY DIFFUSE GALAXIES IN THE COMA CLUSTER},
journal = {The Astrophysical Journal Letters}
}

@article{central,
doi = {10.1088/0004-637X/794/1/59},
url = {https://doi.org/10.1088/0004-637X/794/1/59},
year = {2014},
month = {sep},
publisher = {The American Astronomical Society},
volume = {794},
number = {1},
pages = {59},
author = {Kafle, Prajwal Raj and Sharma, Sanjib and Lewis, Geraint F. and Bland-Hawthorn, Joss},
title = {ON THE SHOULDERS OF GIANTS: PROPERTIES OF THE STELLAR HALO AND THE MILKY WAY MASS DISTRIBUTION},
journal = {The Astrophysical Journal}
}

@ARTICLE{shadow,
       author = {{Falcke}, Heino and {Melia}, Fulvio and {Agol}, Eric},
        title = "{Viewing the Shadow of the Black Hole at the Galactic Center}",
      journal = {Ap. J. Letters},
         year = 2000,
        month = jan,
       volume = {528},
       number = {1},
        pages = {L13-L16},
          doi = {10.1086/312423},
archivePrefix = {arXiv},
       eprint = {astro-ph/9912263},
 primaryClass = {astro-ph},
       adsurl = {https://ui.adsabs.harvard.edu/abs/2000ApJ...528L..13F}
}

@article{shadow2,
   author = "Melia, Fulvio and Falcke, Heino",
   title = "The Supermassive Black Hole at the Galactic Center", 
   journal= "Annual Review of Astronomy and Astrophysics",
   year = "2001",
   volume = "39",
   number = "Volume 39, 2001",
   pages = "309-352",
   doi = "https://doi.org/10.1146/annurev.astro.39.1.309",
   url = "https://www.annualreviews.org/content/journals/10.1146/annurev.astro.39.1.309",
   publisher = "Annual Reviews",
   issn = "1545-4282",
   type = "Journal Article",
  }

@article{EHT,
doi = {10.3847/2041-8213/ac6674},
url = {https://doi.org/10.3847/2041-8213/ac6674},
year = {2022},
month = {may},
publisher = {The American Astronomical Society},
volume = {930},
number = {2},
pages = {L12},
author = {Event Horizon Telescope Collaboration and Akiyama, Kazunori and Alberdi, Antxon and Alef, Walter and Algaba, Juan Carlos and Anantua, Richard and Asada, Keiichi and Azulay, Rebecca and Bach, Uwe and Baczko, Anne-Kathrin and Ball, David and Baloković, Mislav and Barrett, John and Bauböck, Michi and Benson, Bradford A. and Bintley, Dan and Blackburn, Lindy and Blundell, Raymond and Bouman, Katherine L. and Bower, Geoffrey C. and Boyce, Hope and Bremer, Michael and Brinkerink, Christiaan D. and Brissenden, Roger and Britzen, Silke and Broderick, Avery E. and Broguiere, Dominique and Bronzwaer, Thomas and Bustamante, Sandra and Byun, Do-Young and Carlstrom, John E. and Ceccobello, Chiara and Chael, Andrew and Chan, Chi-kwan and Chatterjee, Koushik and Chatterjee, Shami and Chen, Ming-Tang and Chen, Yongjun and Cheng, Xiaopeng and Cho, Ilje and Christian, Pierre and Conroy, Nicholas S. and Conway, John E. and Cordes, James M. and Crawford, Thomas M. and Crew, Geoffrey B. and Cruz-Osorio, Alejandro and Cui, Yuzhu and Davelaar, Jordy and Laurentis, Mariafelicia De and Deane, Roger and Dempsey, Jessica and Desvignes, Gregory and Dexter, Jason and Dhruv, Vedant and Doeleman, Sheperd S. and Dougal, Sean and Dzib, Sergio A. and Eatough, Ralph P. and Emami, Razieh and Falcke, Heino and Farah, Joseph and Fish, Vincent L. and Fomalont, Ed and Ford, H. Alyson and Fraga-Encinas, Raquel and Freeman, William T. and Friberg, Per and Fromm, Christian M. and Fuentes, Antonio and Galison, Peter and Gammie, Charles F. and García, Roberto and Gentaz, Olivier and Georgiev, Boris and Goddi, Ciriaco and Gold, Roman and Gómez-Ruiz, Arturo I. and Gómez, José L. and Gu, Minfeng and Gurwell, Mark and Hada, Kazuhiro and Haggard, Daryl and Haworth, Kari and Hecht, Michael H. and Hesper, Ronald and Heumann, Dirk and Ho, Luis C. and Ho, Paul and Honma, Mareki and Huang, Chih-Wei L. and Huang, Lei and Hughes, David H. and Ikeda, Shiro and Impellizzeri, C. M. Violette and Inoue, Makoto and Issaoun, Sara and James, David J. and Jannuzi, Buell T. and Janssen, Michael and Jeter, Britton and Jiang, Wu and Jiménez-Rosales, Alejandra and Johnson, Michael D. and Jorstad, Svetlana and Joshi, Abhishek V. and Jung, Taehyun and Karami, Mansour and Karuppusamy, Ramesh and Kawashima, Tomohisa and Keating, Garrett K. and Kettenis, Mark and Kim, Dong-Jin and Kim, Jae-Young and Kim, Jongsoo and Kim, Junhan and Kino, Motoki and Koay, Jun Yi and Kocherlakota, Prashant and Kofuji, Yutaro and Koch, Patrick M. and Koyama, Shoko and Kramer, Carsten and Kramer, Michael and Krichbaum, Thomas P. and Kuo, Cheng-Yu and Bella, Noemi La and Lauer, Tod R. and Lee, Daeyoung and Lee, Sang-Sung and Leung, Po Kin and Levis, Aviad and Li, Zhiyuan and Lico, Rocco and Lindahl, Greg and Lindqvist, Michael and Lisakov, Mikhail and Liu, Jun and Liu, Kuo and Liuzzo, Elisabetta and Lo, Wen-Ping and Lobanov, Andrei P. and Loinard, Laurent and Lonsdale, Colin J. and Lu, Ru-Sen and Mao, Jirong and Marchili, Nicola and Markoff, Sera and Marrone, Daniel P. and Marscher, Alan P. and Martí-Vidal, Iván and Matsushita, Satoki and Matthews, Lynn D. and Medeiros, Lia and Menten, Karl M. and Michalik, Daniel and Mizuno, Izumi and Mizuno, Yosuke and Moran, James M. and Moriyama, Kotaro and Moscibrodzka, Monika and Müller, Cornelia and Mus, Alejandro and Musoke, Gibwa and Myserlis, Ioannis and Nadolski, Andrew and Nagai, Hiroshi and Nagar, Neil M. and Nakamura, Masanori and Narayan, Ramesh and Narayanan, Gopal and Natarajan, Iniyan and Nathanail, Antonios and Fuentes, Santiago Navarro and Neilsen, Joey and Neri, Roberto and Ni, Chunchong and Noutsos, Aristeidis and Nowak, Michael A. and Oh, Junghwan and Okino, Hiroki and Olivares, Héctor and Ortiz-León, Gisela N. and Oyama, Tomoaki and Özel, Feryal and Palumbo, Daniel C. M. and Paraschos, Georgios Filippos and Park, Jongho and Parsons, Harriet and Patel, Nimesh and Pen, Ue-Li and Pesce, Dominic W. and Piétu, Vincent and Plambeck, Richard and PopStefanija, Aleksandar and Porth, Oliver and Pötzl, Felix M. and Prather, Ben and Preciado-López, Jorge A. and Psaltis, Dimitrios and Pu, Hung-Yi and Ramakrishnan, Venkatessh and Rao, Ramprasad and Rawlings, Mark G. and Raymond, Alexander W. and Rezzolla, Luciano and Ricarte, Angelo and Ripperda, Bart and Roelofs, Freek and Rogers, Alan and Ros, Eduardo and Romero-Cañizales, Cristina and Roshanineshat, Arash and Rottmann, Helge and Roy, Alan L. and Ruiz, Ignacio and Ruszczyk, Chet and Rygl, Kazi L. J. and Sánchez, Salvador and Sánchez-Argüelles, David and Sánchez-Portal, Miguel and Sasada, Mahito and Satapathy, Kaushik and Savolainen, Tuomas and Schloerb, F. Peter and Schonfeld, Jonathan and Schuster, Karl-Friedrich and Shao, Lijing and Shen, Zhiqiang and Small, Des and Sohn, Bong Won and SooHoo, Jason and Souccar, Kamal and Sun, He and Tazaki, Fumie and Tetarenko, Alexandra J. and Tiede, Paul and Tilanus, Remo P. J. and Titus, Michael and Torne, Pablo and Traianou, Efthalia and Trent, Tyler and Trippe, Sascha and Turk, Matthew and van Bemmel, Ilse and van Langevelde, Huib Jan and van Rossum, Daniel R. and Vos, Jesse and Wagner, Jan and Ward-Thompson, Derek and Wardle, John and Weintroub, Jonathan and Wex, Norbert and Wharton, Robert and Wielgus, Maciek and Wiik, Kaj and Witzel, Gunther and Wondrak, Michael F. and Wong, George N. and Wu, Qingwen and Yamaguchi, Paul and Yoon, Doosoo and Young, André and Young, Ken and Younsi, Ziri and Yuan, Feng and Yuan, Ye-Fei and Zensus, J. Anton and Zhang, Shuo and Zhao, Guang-Yao and Zhao, Shan-Shan and Agurto, Claudio and Allardi, Alexander and Amestica, Rodrigo and Araneda, Juan Pablo and Arriagada, Oriel and Berghuis, Jennie L. and Bertarini, Alessandra and Berthold, Ryan and Blanchard, Jay and Brown, Ken and Cárdenas, Mauricio and Cantzler, Michael and Caro, Patricio and Castillo-Domínguez, Edgar and Chan, Tin Lok and Chang, Chih-Cheng and Chang, Dominic O. and Chang, Shu-Hao and Chang, Song-Chu and Chen, Chung-Chen and Chilson, Ryan and Chuter, Tim C. and Ciechanowicz, Miroslaw and Colin-Beltran, Edgar and Coulson, Iain M. and Crowley, Joseph and Degenaar, Nathalie and Dornbusch, Sven and Durán, Carlos A. and Everett, Wendeline B. and Faber, Aaron and Forster, Karl and Fuchs, Miriam M. and Gale, David M. and Geertsema, Gertie and González, Edouard and Graham, Dave and Gueth, Frédéric and Halverson, Nils W. and Han, Chih-Chiang and Han, Kuo-Chang and Hasegawa, Yutaka and Hernández-Rebollar, José Luis and Herrera, Cristian and Herrero-Illana, Ruben and Heyminck, Stefan and Hirota, Akihiko and Hoge, James and Hostler Schimpf, Shelbi R. and Howie, Ryan E. and Huang, Yau-De and Jiang, Homin and Jinchi, Hao and John, David and Kimura, Kimihiro and Klein, Thomas and Kubo, Derek and Kuroda, John and Kwon, Caleb and Lacasse, Richard and Laing, Robert and Leitch, Erik M. and Li, Chao-Te and Liu, Ching-Tang and Liu, Kuan-Yu and Lin, Lupin C.-C. and Lu, Li-Ming and Mac-Auliffe, Felipe and Martin-Cocher, Pierre and Matulonis, Callie and Maute, John K. and Messias, Hugo and Meyer-Zhao, Zheng and Montaña, Alfredo and Montenegro-Montes, Francisco and Montgomerie, William and Moreno Nolasco, Marcos Emir and Muders, Dirk and Nishioka, Hiroaki and Norton, Timothy J. and Nystrom, George and Ogawa, Hideo and Olivares, Rodrigo and Oshiro, Peter and Pérez-Beaupuits, Juan Pablo and Parra, Rodrigo and Phillips, Neil M. and Poirier, Michael and Pradel, Nicolas and Qiu, Richard and Raffin, Philippe A. and Rahlin, Alexandra S. and Ramírez, Jorge and Ressler, Sean and Reynolds, Mark and Rodríguez-Montoya, Iván and Saez-Madain, Alejandro F. and Santana, Jorge and Shaw, Paul and Shirkey, Leslie E. and Silva, Kevin M. and Snow, William and Sousa, Don and Sridharan, T. K. and Stahm, William and Stark, Anthony A. and Test, John and Torstensson, Karl and Venegas, Paulina and Walther, Craig and Wei, Ta-Shun and White, Chris and Wieching, Gundolf and Wijnands, Rudy and Wouterloot, Jan G. A. and Yu, Chen-Yu and Yu (于威), Wei and Zeballos, Milagros},
title = {First Sagittarius A* Event Horizon Telescope Results. I. The Shadow of the Supermassive Black Hole in the Center of the Milky Way},
journal = {The Astrophysical Journal Letters}
}

@ARTICLE{EHT2,
       author = {{Event Horizon Telescope Collaboration} and {Akiyama}, Kazunori and {Alberdi}, Antxon and {Alef}, Walter and {Algaba}, Juan Carlos and {Anantua}, Richard and {Asada}, Keiichi and {Azulay}, Rebecca and {Bach}, Uwe and {Baczko}, Anne-Kathrin and {Ball}, David and {Balokovi{\'c}}, Mislav and {Bandyopadhyay}, Bidisha and {Barrett}, John and {Baub{\"o}ck}, Michi and {Benson}, Bradford A. and {Bintley}, Dan and {Blackburn}, Lindy and {Blundell}, Raymond and {Bouman}, Katherine L. and {Bower}, Geoffrey C. and {Boyce}, Hope and {Bremer}, Michael and {Brissenden}, Roger and {Britzen}, Silke and {Broderick}, Avery E. and {Broguiere}, Dominique and {Bronzwaer}, Thomas and {Bustamante}, Sandra and {Carlstrom}, John E. and {Chael}, Andrew and {Chan}, Chi-kwan and {Chang}, Dominic O. and {Chatterjee}, Koushik and {Chatterjee}, Shami and {Chen}, Ming-Tang and {Chen}, Yongjun and {Cheng}, Xiaopeng and {Cho}, Ilje and {Christian}, Pierre and {Conroy}, Nicholas S. and {Conway}, John E. and {Crawford}, Thomas M. and {Crew}, Geoffrey B. and {Cruz-Osorio}, Alejandro and {Cui}, Yuzhu and {Dahale}, Rohan and {Davelaar}, Jordy and {De Laurentis}, Mariafelicia and {Deane}, Roger and {Dempsey}, Jessica and {Desvignes}, Gregory and {Dexter}, Jason and {Dhruv}, Vedant and {Dihingia}, Indu K. and {Doeleman}, Sheperd S. and {Dzib}, Sergio A. and {Eatough}, Ralph P. and {Emami}, Razieh and {Falcke}, Heino and {Farah}, Joseph and {Fish}, Vincent L. and {Fomalont}, Edward and {Ford}, H. Alyson and {Foschi}, Marianna and {Fraga-Encinas}, Raquel and {Freeman}, William T. and {Friberg}, Per and {Fromm}, Christian M. and {Fuentes}, Antonio and {Galison}, Peter and {Gammie}, Charles F. and {Garc{\'\i}a}, Roberto and {Gentaz}, Olivier and {Georgiev}, Boris and {Goddi}, Ciriaco and {Gold}, Roman and {G{\'o}mez-Ruiz}, Arturo I. and {G{\'o}mez}, Jos{\'e} L. and {Gu}, Minfeng and {Gurwell}, Mark and {Hada}, Kazuhiro and {Haggard}, Daryl and {Hesper}, Ronald and {Heumann}, Dirk and {Ho}, Luis C. and {Ho}, Paul and {Honma}, Mareki and {Huang}, Chih-Wei L. and {Huang}, Lei and {Hughes}, David H. and {Ikeda}, Shiro and {Violette Impellizzeri}, C.~M. and {Inoue}, Makoto and {Issaoun}, Sara and {James}, David J. and {Jannuzi}, Buell T. and {Janssen}, Michael and {Jeter}, Britton and {Jiang}, Wu and {Jim{\'e}nez-Rosales}, Alejandra and {Johnson}, Michael D. and {Jorstad}, Svetlana and {Jones}, Adam C. and {Joshi}, Abhishek V. and {Jung}, Taehyun and {Karuppusamy}, Ramesh and {Kawashima}, Tomohisa and {Keating}, Garrett K. and {Kettenis}, Mark and {Kim}, Dong-Jin and {Kim}, Jae-Young and {Kim}, Jongsoo and {Kim}, Junhan and {Kino}, Motoki and {Koay}, Jun Yi and {Kocherlakota}, Prashant and {Kofuji}, Yutaro and {Koch}, Patrick M. and {Koyama}, Shoko and {Kramer}, Carsten and {Kramer}, Joana A. and {Kramer}, Michael and {Krichbaum}, Thomas P. and {Kuo}, Cheng-Yu and {La Bella}, Noemi and {Lee}, Sang-Sung and {Levis}, Aviad and {Li}, Zhiyuan and {Lico}, Rocco and {Lindahl}, Greg and {Lindqvist}, Michael and {Lisakov}, Mikhail and {Liu}, Jun and {Liu}, Kuo and {Liuzzo}, Elisabetta and {Lo}, Wen-Ping and {Lobanov}, Andrei P. and {Loinard}, Laurent and {Lonsdale}, Colin J. and {Lowitz}, Amy E. and {Lu}, Ru-Sen and {MacDonald}, Nicholas R. and {Mao}, Jirong and {Marchili}, Nicola and {Markoff}, Sera and {Marrone}, Daniel P. and {Marscher}, Alan P. and {Mart{\'\i}-Vidal}, Iv{\'a}n and {Matsushita}, Satoki and {Matthews}, Lynn D. and {Medeiros}, Lia and {Menten}, Karl M. and {Mizuno}, Izumi and {Mizuno}, Yosuke and {Montgomery}, Joshua and {Moran}, James M. and {Moriyama}, Kotaro and {Moscibrodzka}, Monika and {Mulaudzi}, Wanga and {M{\"u}ller}, Cornelia and {M{\"u}ller}, Hendrik and {Mus}, Alejandro and {Musoke}, Gibwa and {Myserlis}, Ioannis and {Nagai}, Hiroshi and {Nagar}, Neil M. and {Nakamura}, Masanori and {Narayanan}, Gopal and {Natarajan}, Iniyan and {Nathanail}, Antonios and {Fuentes}, Santiago Navarro and {Neilsen}, Joey and {Ni}, Chunchong and {Nowak}, Michael A. and {Oh}, Junghwan and {Okino}, Hiroki and {Olivares}, H{\'e}ctor and {Oyama}, Tomoaki and {{\"O}zel}, Feryal and {Palumbo}, Daniel C.~M. and {Paraschos}, Georgios Filippos and {Park}, Jongho and {Parsons}, Harriet and {Patel}, Nimesh and {Pen}, Ue-Li and {Pesce}, Dominic W. and {Pi{\'e}tu}, Vincent and {PopStefanija}, Aleksandar and {Porth}, Oliver and {Prather}, Ben and {Psaltis}, Dimitrios and {Pu}, Hung-Yi and {Ramakrishnan}, Venkatessh and {Rao}, Ramprasad and {Rawlings}, Mark G. and {Raymond}, Alexander W. and {Rezzolla}, Luciano and {Ricarte}, Angelo and {Ripperda}, Bart},
        title = "{The persistent shadow of the supermassive black hole of M 87. I. Observations, calibration, imaging, and analysis}",
      journal = {A\&A},
         year = 2024,
        month = jan,
       volume = {681},
          eid = {A79},
        pages = {A79},
          doi = {10.1051/0004-6361/202347932},
       adsurl = {https://ui.adsabs.harvard.edu/abs/2024A&A...681A..79E}
}

@ARTICLE{EHT3,
       author = {{Event Horizon Telescope Collaboration} and {Akiyama}, Kazunori and {Albentosa-Ru{\'\i}z}, Ezequiel and {Alberdi}, Antxon and {Alef}, Walter and {Algaba}, Juan Carlos and {Anantua}, Richard and {Asada}, Keiichi and {Azulay}, Rebecca and {Bach}, Uwe and {Baczko}, Anne-Kathrin and {Ball}, David and {Balokovi{\'c}}, Mislav and {Bandyopadhyay}, Bidisha and {Barrett}, John and {Baub{\"o}ck}, Michi and {Benson}, Bradford A. and {Bintley}, Dan and {Blackburn}, Lindy and {Blundell}, Raymond and {Bouman}, Katherine L. and {Bower}, Geoffrey C. and {Bremer}, Michael and {Brissenden}, Roger and {Britzen}, Silke and {Broderick}, Avery E. and {Broguiere}, Dominique and {Bronzwaer}, Thomas and {Bustamante}, Sandra and {Carlstrom}, John E. and {Chael}, Andrew and {Chan}, Chi-kwan and {Chang}, Dominic O. and {Chatterjee}, Koushik and {Chatterjee}, Shami and {Chen}, Ming-Tang and {Chen}, Yongjun and {Cheng}, Xiaopeng and {Cho}, Ilje and {Christian}, Pierre and {Conroy}, Nicholas S. and {Conway}, John E. and {Crawford}, Thomas M. and {Crew}, Geoffrey B. and {Cruz-Osorio}, Alejandro and {Cui}, Yuzhu and {Curd}, Brandon and {Dahale}, Rohan and {Davelaar}, Jordy and {De Laurentis}, Mariafelicia and {Deane}, Roger and {Dempsey}, Jessica and {Desvignes}, Gregory and {Dexter}, Jason and {Dhruv}, Vedant and {Dihingia}, Indu K. and {Doeleman}, Sheperd S. and {Dzib}, Sergio A. and {Eatough}, Ralph P. and {Emami}, Razieh and {Falcke}, Heino and {Farah}, Joseph and {Fish}, Vincent L. and {Fomalont}, Edward and {Ford}, H. Alyson and {Foschi}, Marianna and {Fraga-Encinas}, Raquel and {Freeman}, William T. and {Friberg}, Per and {Fromm}, Christian M. and {Fuentes}, Antonio and {Galison}, Peter and {Gammie}, Charles F. and {Garc{\'\i}a}, Roberto and {Gentaz}, Olivier and {Georgiev}, Boris and {Goddi}, Ciriaco and {Gold}, Roman and {G{\'o}mez-Ruiz}, Arturo I. and {G{\'o}mez}, Jos{\'e} L. and {Gu}, Minfeng and {Gurwell}, Mark and {Hada}, Kazuhiro and {Haggard}, Daryl and {Hesper}, Ronald and {Heumann}, Dirk and {Ho}, Luis C. and {Ho}, Paul and {Honma}, Mareki and {Huang}, Chih-Wei L. and {Huang}, Lei and {Hughes}, David H. and {Ikeda}, Shiro and {Impellizzeri}, C.~M. Violette and {Inoue}, Makoto and {Issaoun}, Sara and {James}, David J. and {Jannuzi}, Buell T. and {Janssen}, Michael and {Jeter}, Britton and {Jiang}, Wu and {Jim{\'e}nez-Rosales}, Alejandra and {Johnson}, Michael D. and {Jorstad}, Svetlana and {Jones}, Adam C. and {Joshi}, Abhishek V. and {Jung}, Taehyun and {Karuppusamy}, Ramesh and {Kawashima}, Tomohisa and {Keating}, Garrett K. and {Kettenis}, Mark and {Kim}, Dong-Jin and {Kim}, Jae-Young and {Kim}, Jongsoo and {Kim}, Junhan and {Kino}, Motoki and {Koay}, Jun Yi and {Kocherlakota}, Prashant and {Kofuji}, Yutaro and {Koch}, Patrick M. and {Koyama}, Shoko and {Kramer}, Carsten and {Kramer}, Joana A. and {Kramer}, Michael and {Krichbaum}, Thomas P. and {Kuo}, Cheng-Yu and {La Bella}, Noemi and {Lee}, Sang-Sung and {Levis}, Aviad and {Li}, Zhiyuan and {Lico}, Rocco and {Lindahl}, Greg and {Lindqvist}, Michael and {Lisakov}, Mikhail and {Liu}, Jun and {Liu}, Kuo and {Liuzzo}, Elisabetta and {Lo}, Wen-Ping and {Lobanov}, Andrei P. and {Loinard}, Laurent and {Lonsdale}, Colin J. and {Lowitz}, Amy E. and {Lu}, Ru-Sen and {MacDonald}, Nicholas R. and {Mao}, Jirong and {Marchili}, Nicola and {Markoff}, Sera and {Marrone}, Daniel P. and {Marscher}, Alan P. and {Mart{\'\i}-Vidal}, Iv{\'a}n and {Matsushita}, Satoki and {Matthews}, Lynn D. and {Medeiros}, Lia and {Menten}, Karl M. and {Mizuno}, Izumi and {Mizuno}, Yosuke and {Montgomery}, Joshua and {Moran}, James M. and {Moriyama}, Kotaro and {Moscibrodzka}, Monika and {Mulaudzi}, Wanga and {M{\"u}ller}, Cornelia and {M{\"u}ller}, Hendrik and {Mus}, Alejandro and {Musoke}, Gibwa and {Myserlis}, Ioannis and {Nagai}, Hiroshi and {Nagar}, Neil M. and {Nair}, Dhanya G. and {Nakamura}, Masanori and {Narayanan}, Gopal and {Natarajan}, Iniyan and {Nathanail}, Antonios and {Navarro Fuentes}, Santiago and {Neilsen}, Joey and {Ni}, Chunchong and {Nowak}, Michael A. and {Oh}, Junghwan and {Okino}, Hiroki and {Ra{\'u}l Olivares S{\'a}nchez}, H{\'e}ctor and {Oyama}, Tomoaki and {{\"O}zel}, Feryal and {Palumbo}, Daniel C.~M. and {Paraschos}, Georgios Filippos and {Park}, Jongho and {Parsons}, Harriet and {Patel}, Nimesh and {Pen}, Ue-Li and {Pesce}, Dominic W. and {Pi{\'e}tu}, Vincent and {PopStefanija}, Aleksandar and {Porth}, Oliver and {Prather}, Ben and {Principe}, Giacomo and {Psaltis}, Dimitrios and {Pu}, Hung-Yi and {Ramakrishnan}, Venkatessh and {Rao}, Ramprasad and {Rawlings}, Mark G. and {Rezzolla}, Luciano},
        title = "{The persistent shadow of the supermassive black hole of M87: II. Model comparisons and theoretical interpretations}",
      journal = {A\&A},
         year = 2025,
        month = jan,
       volume = {693},
          eid = {A265},
        pages = {A265},
          doi = {10.1051/0004-6361/202451296},
       adsurl = {https://ui.adsabs.harvard.edu/abs/2025A&A...693A.265E}
}

@book{BH,
  title={Black hole physics: Basic concepts and new developments},
  author={Frolov, Valeri and Novikov, Igor},
  volume={96},
  year={2012},
  publisher={Springer Science \& Business Media}
}

@ARTICLE{shadow4,
       author = {{Perlick}, Volker and {Tsupko}, Oleg Yu.},
        title = "{Calculating black hole shadows: Review of analytical studies}",
      journal = {Physical Reports},
         year = 2022,
        month = feb,
       volume = {947},
        pages = {1-39},
          doi = {10.1016/j.physrep.2021.10.004},
archivePrefix = {arXiv},
       eprint = {2105.07101},
 primaryClass = {gr-qc},
       adsurl = {https://ui.adsabs.harvard.edu/abs/2022PhR...947....1P}
}

@article{sstar,
  title = {Probing dense environments around Sgr A* with S-star dynamics},
  author = {Tomaselli, Giovanni Maria and Caputo, Andrea},
  journal = {Phys. Rev. D},
  volume = {113},
  issue = {8},
  pages = {083035},
  numpages = {18},
  year = {2026},
  month = {Apr},
  publisher = {American Physical Society},
  doi = {10.1103/pm7p-c53w},
  url = {https://link.aps.org/doi/10.1103/pm7p-c53w}
}

@ARTICLE{sstar2,
       author = {{Gravity Collaboration} and {Abd El Dayem}, K. and {Abuter}, R. and {Aimar}, N. and {Amaro Seoane}, P. and {Amorim}, A. and {Beck}, J. and {Berger}, J.~P. and {Bonnet}, H. and {Bourdarot}, G. and {Brandner}, W. and {Cardoso}, V. and {Capuzzo Dolcetta}, R. and {Cl{\'e}net}, Y. and {Davies}, R. and {de Zeeuw}, P.~T. and {Drescher}, A. and {Eckart}, A. and {Eisenhauer}, F. and {Feuchtgruber}, H. and {Finger}, G. and {F{\"o}rster Schreiber}, N.~M. and {Foschi}, A. and {Gao}, F. and {Garcia}, P. and {Gendron}, E. and {Genzel}, R. and {Gillessen}, S. and {Hartl}, M. and {Haubois}, X. and {Haussmann}, F. and {Hei{\ss}el}, G. and {Henning}, T. and {Hippler}, S. and {Horrobin}, M. and {Jochum}, L. and {Jocou}, L. and {Kaufer}, A. and {Kervella}, P. and {Lacour}, S. and {Lapeyr{\`e}re}, V. and {Le Bouquin}, J.-B. and {L{\'e}na}, P. and {Lutz}, D. and {Mang}, F. and {More}, N. and {Ott}, T. and {Paumard}, T. and {Perraut}, K. and {Perrin}, G. and {Pfuhl}, O. and {Rabien}, S. and {Ribeiro}, D.~C. and {Sadun Bordoni}, M. and {Scheithauer}, S. and {Shangguan}, J. and {Shimizu}, T. and {Stadler}, J. and {Straub}, O. and {Straubmeier}, C. and {Sturm}, E. and {Tacconi}, L.~J. and {Urso}, I. and {Vincent}, F. and {von Fellenberg}, S.~D. and {Widmann}, F. and {Wieprecht}, E. and {Woillez}, J. and {Zhang}, F.},
        title = "{Improving constraints on the extended mass distribution in the Galactic center with stellar orbits}",
      journal = {A\&A},
         year = 2024,
        month = dec,
       volume = {692},
          eid = {A242},
        pages = {A242},
          doi = {10.1051/0004-6361/202452274},
archivePrefix = {arXiv},
       eprint = {2409.12261},
 primaryClass = {astro-ph.GA},
       adsurl = {https://ui.adsabs.harvard.edu/abs/2024A&A...692A.242G}
}

@article{sstar3,
doi = {10.3847/1538-4357/ab9c1c},
url = {https://doi.org/10.3847/1538-4357/ab9c1c},
year = {2020},
month = {aug},
publisher = {The American Astronomical Society},
volume = {899},
number = {1},
pages = {50},
author = {Peissker, Florian and Eckart, Andreas and Zaja\v{c}ek, Michal and Ali, Basel and Parsa, Marzieh},
title = {S62 and S4711: Indications of a Population of Faint Fast-moving Stars inside the S2 Orbit—S4711 on a 7.6 yr Orbit around Sgr A*},
journal = {The Astrophysical Journal}
}

@article{sstar4,
doi = {10.1088/0264-9381/33/11/113001},
url = {https://doi.org/10.1088/0264-9381/33/11/113001},
year = {2016},
month = {may},
publisher = {IOP Publishing},
volume = {33},
number = {11},
pages = {113001},
author = {Johannsen, Tim},
title = {Sgr A* and general relativity},
journal = {Classical and Quantum Gravity}
}

@article{sstar5,
    author = {Navarrete, C'\esar and Vázquez-Ch\'avez, Fernando and Cruz-Osorio, Alejandro and Ortiz, Néstor},
    title = {Testing black hole space–times with the S2 star orbit: a Bayesian comparison},
    journal = {Monthly Notices of the Royal Astronomical Society},
    volume = {546},
    number = {3},
    pages = {stag059},
    year = {2026},
    month = {03},
    issn = {0035-8711},
    doi = {10.1093/mnras/stag059},
    url = {https://doi.org/10.1093/mnras/stag059},
    eprint = {https://academic.oup.com/mnras/article-pdf/546/3/stag059/66339812/stag059.pdf},
}

@article{sstar6,
	author = {{Hei\ss{}el, G.} and {Paumard, T.} and {Perrin, G.} and {Vincent, F.}},
	title = {The dark mass signature in the orbit of S2},
	DOI= "10.1051/0004-6361/202142114",
	url= "https://doi.org/10.1051/0004-6361/202142114",
	journal = {A\&A},
	year = 2022,
	volume = 660,
	pages = "A13",
}

@ARTICLE{oort,
       author = {{Oort}, J.~H.},
        title = "{The force exerted by the stellar system in the direction perpendicular to the galactic plane and some related problems}",
      journal = {Bulletin of the Astronomical Institutes of the Netherlands},
         year = 1932,
        month = aug,
       volume = {6},
        pages = {249},
       adsurl = {https://ui.adsabs.harvard.edu/abs/1932BAN.....6..249O}
}

@ARTICLE{zwicky,
       author = {{Zwicky}, F.},
        title = "{On the Masses of Nebulae and of Clusters of Nebulae}",
      journal = {Ap. J.},
         year = 1937,
        month = oct,
       volume = {86},
        pages = {217},
          doi = {10.1086/143864},
       adsurl = {https://ui.adsabs.harvard.edu/abs/1937ApJ....86..217Z}
}

@article{snae,
  title = {Nobel Lecture: Accelerating expansion of the Universe through observations of distant supernovae},
  author = {Schmidt, Brian P.},
  journal = {Rev. Mod. Phys.},
  volume = {84},
  issue = {3},
  pages = {1151--1163},
  numpages = {0},
  year = {2012},
  month = {Aug},
  publisher = {American Physical Society},
  doi = {10.1103/RevModPhys.84.1151},
  url = {https://link.aps.org/doi/10.1103/RevModPhys.84.1151}
}

@article{planck,
	author = {{Planck Collaboration} and {Aghanim, N.} and {Akrami, Y.} and {Arroja, F.} and {Ashdown, M.} and {Aumont, J.} and {Baccigalupi, C.} and {Ballardini, M.} and {Banday, A. J.} and {Barreiro, R. B.} and {Bartolo, N.} and {Basak, S.} and {Battye, R.} and {Benabed, K.} and {Bernard, J.-P.} and {Bersanelli, M.} and {Bielewicz, P.} and {Bock, J. J.} and {Bond, J. R.} and {Borrill, J.} and {Bouchet, F. R.} and {Boulanger, F.} and {Bucher, M.} and {Burigana, C.} and {Butler, R. C.} and {Calabrese, E.} and {Cardoso, J.-F.} and {Carron, J.} and {Casaponsa, B.} and {Challinor, A.} and {Chiang, H. C.} and {Colombo, L. P. L.} and {Combet, C.} and {Contreras, D.} and {Crill, B. P.} and {Cuttaia, F.} and {de Bernardis, P.} and {de Zotti, G.} and {Delabrouille, J.} and {Delouis, J.-M.} and {D\'esert, F.-X.} and {Di Valentino, E.} and {Dickinson, C.} and {Diego, J. M.} and {Donzelli, S.} and {Dor\'e, O.} and {Douspis, M.} and {Ducout, A.} and {Dupac, X.} and {Efstathiou, G.} and {Elsner, F.} and {En\ss{}lin, T. A.} and {Eriksen, H. K.} and {Falgarone, E.} and {Fantaye, Y.} and {Fergusson, J.} and {Fernandez-Cobos, R.} and {Finelli, F.} and {Forastieri, F.} and {Frailis, M.} and {Franceschi, E.} and {Frolov, A.} and {Galeotta, S.} and {Galli, S.} and {Ganga, K.} and {G\'enova-Santos, R. T.} and {Gerbino, M.} and {Ghosh, T.} and {Gonz\'alez-Nuevo, J.} and {G\'orski, K. M.} and {Gratton, S.} and {Gruppuso, A.} and {Gudmundsson, J. E.} and {Hamann, J.} and {Handley, W.} and {Hansen, F. K.} and {Helou, G.} and {Herranz, D.} and {Hildebrandt, S. R.} and {Hivon, E.} and {Huang, Z.} and {Jaffe, A. H.} and {Jones, W. C.} and {Karakci, A.} and {Keih\"anen, E.} and {Keskitalo, R.} and {Kiiveri, K.} and {Kim, J.} and {Kisner, T. S.} and {Knox, L.} and {Krachmalnicoff, N.} and {Kunz, M.} and {Kurki-Suonio, H.} and {Lagache, G.} and {Lamarre, J.-M.} and {Langer, M.} and {Lasenby, A.} and {Lattanzi, M.} and {Lawrence, C. R.} and {Le Jeune, M.} and {Leahy, J. P.} and {Lesgourgues, J.} and {Levrier, F.} and {Lewis, A.} and {Liguori, M.} and {Lilje, P. B.} and {Lilley, M.} and {Lindholm, V.} and {L\'opez-Caniego, M.} and {Lubin, P. M.} and {Ma, Y.-Z.} and {Mac\'{\i}as-P\'erez, J. F.} and {Maggio, G.} and {Maino, D.} and {Mandolesi, N.} and {Mangilli, A.} and {Marcos-Caballero, A.} and {Maris, M.} and {Martin, P. G.} and {Martinelli, M.} and {Mart\'{\i}nez-Gonz\'alez, E.} and {Matarrese, S.} and {Mauri, N.} and {McEwen, J. D.} and {Meerburg, P. D.} and {Meinhold, P. R.} and {Melchiorri, A.} and {Mennella, A.} and {Migliaccio, M.} and {Millea, M.} and {Mitra, S.} and {Miville-Desch\^enes, M.-A.} and {Molinari, D.} and {Moneti, A.} and {Montier, L.} and {Morgante, G.} and {Moss, A.} and {Mottet, S.} and {M\"unchmeyer, M.} and {Natoli, P.} and {N\o{}rgaard-Nielsen, H. U.} and {Oxborrow, C. A.} and {Pagano, L.} and {Paoletti, D.} and {Partridge, B.} and {Patanchon, G.} and {Pearson, T. J.} and {Peel, M.} and {Peiris, H. V.} and {Perrotta, F.} and {Pettorino, V.} and {Piacentini, F.} and {Polastri, L.} and {Polenta, G.} and {Puget, J.-L.} and {Rachen, J. P.} and {Reinecke, M.} and {Remazeilles, M.} and {Renault, C.} and {Renzi, A.} and {Rocha, G.} and {Rosset, C.} and {Roudier, G.} and {Rubi\~no-Mart\'{\i}n, J. A.} and {Ruiz-Granados, B.} and {Salvati, L.} and {Sandri, M.} and {Savelainen, M.} and {Scott, D.} and {Shellard, E. P. S.} and {Shiraishi, M.} and {Sirignano, C.} and {Sirri, G.} and {Spencer, L. D.} and {Sunyaev, R.} and {Suur-Uski, A.-S.} and {Tauber, J. A.} and {Tavagnacco, D.} and {Tenti, M.} and {Terenzi, L.} and {Toffolatti, L.} and {Tomasi, M.} and {Trombetti, T.} and {Valiviita, J.} and {Van Tent, B.} and {Vibert, L.} and {Vielva, P.} and {Villa, F.} and {Vittorio, N.} and {Wandelt, B. D.} and {Wehus, I. K.} and {White, M.} and {White, S. D. M.} and {Zacchei, A.} and {Zonca, A.}},
	title = {Planck 2018 results - I. Overview and the cosmological legacy of Planck},
	DOI= "10.1051/0004-6361/201833880",
	url= "https://doi.org/10.1051/0004-6361/201833880",
	journal = {A\&A},
	year = 2020,
	volume = 641,
	pages = "A1",
}

@article{wimp,
doi = {10.1088/1361-6633/aab913},
url = {https://doi.org/10.1088/1361-6633/aab913},
year = {2018},
month = {may},
publisher = {IOP Publishing},
volume = {81},
number = {6},
pages = {066201},
author = {Roszkowski, Leszek and Sessolo, Enrico Maria and Trojanowski, Sebastian},
title = {WIMP dark matter candidates and searches—current status and future prospects},
journal = {Reports on Progress in Physics}
}

@book{weinberg,
  title={The quantum theory of fields},
  author={Weinberg, Steven},
  volume={2},
  year={1995},
  publisher={Cambridge university press}
}

@article{axion,
doi = {10.1088/1126-6708/2006/06/051},
url = {https://doi.org/10.1088/1126-6708/2006/06/051},
year = {2006},
month = {jun},
volume = {2006},
number = {06},
pages = {051},
author = {Peter Svrcek and Edward Witten},
title = {Axions in string theory},
journal = {Journal of High Energy Physics}
}

@ARTICLE{ster,
       author = {{Boyarsky}, A. and {Drewes}, M. and {Lasserre}, T. and {Mertens}, S. and {Ruchayskiy}, O.},
        title = "{Sterile neutrino Dark Matter}",
      journal = {Progress in Particle and Nuclear Physics},
         year = 2019,
        month = jan,
       volume = {104},
        pages = {1-45},
          doi = {10.1016/j.ppnp.2018.07.004},
archivePrefix = {arXiv},
       eprint = {1807.07938},
 primaryClass = {hep-ph},
       adsurl = {https://ui.adsabs.harvard.edu/abs/2019PrPNP.104....1B}
}

@article{pbh,
doi = {10.1088/1361-6471/abc534},
url = {https://doi.org/10.1088/1361-6471/abc534},
year = {2021},
month = {feb},
publisher = {IOP Publishing},
volume = {48},
number = {4},
pages = {043001},
author = {Green, Anne M and Kavanagh, Bradley J},
title = {Primordial black holes as a dark matter candidate},
journal = {Journal of Physics G: Nuclear and Particle Physics}
}

@article{search,
doi = {10.1088/1742-6596/2502/1/012004},
url = {https://doi.org/10.1088/1742-6596/2502/1/012004},
year = {2023},
month = {may},
publisher = {IOP Publishing},
volume = {2502},
number = {1},
pages = {012004},
author = {Cebrián, Susana},
title = {Review on dark matter searches},
journal = {Journal of Physics: Conference Series}
}

@ARTICLE{nfw,
       author = {{Navarro}, Julio F. and {Frenk}, Carlos S. and {White}, Simon D.~M.},
        title = "{The Structure of Cold Dark Matter Halos}",
      journal = {Ap. J.},
         year = 1996,
        month = may,
       volume = {462},
        pages = {563},
          doi = {10.1086/177173},
archivePrefix = {arXiv},
       eprint = {astro-ph/9508025},
 primaryClass = {astro-ph},
       adsurl = {https://ui.adsabs.harvard.edu/abs/1996ApJ...462..563N}
}

@article{nfw2,
doi = {10.1086/304888},
url = {https://doi.org/10.1086/304888},
year = {1997},
month = {dec},
publisher = {},
volume = {490},
number = {2},
pages = {493},
author = {Navarro, Julio F. and Frenk, Carlos S. and White, Simon D. M.},
title = {A Universal Density Profile from Hierarchical Clustering},
journal = {The Astrophysical Journal}
}

@article{sat,
doi = {10.3847/1538-4357/abbd92},
url = {https://doi.org/10.3847/1538-4357/abbd92},
year = {2020},
month = {nov},
publisher = {The American Astronomical Society},
volume = {903},
number = {2},
pages = {130},
author = {Seo, Gangil and Sohn, Jubee and Lee, Myung Gyoon},
title = {Tracing Dark Matter Halos with Satellite Kinematics and the Central Stellar Velocity Dispersion of Galaxies},
journal = {The Astrophysical Journal}
}

@article{sat2,
    author = {Lange, Johannes U and van den Bosch, Frank C and Zentner, Andrew R and Wang, Kuan and Villarreal, Antonia Sierra},
    title = {Updated results on the galaxy–halo connection from satellite kinematics in SDSS},
    journal = {Monthly Notices of the Royal Astronomical Society},
    volume = {487},
    number = {3},
    pages = {3112-3129},
    year = {2019},
    month = {08},
    issn = {0035-8711},
    doi = {10.1093/mnras/stz1466},
    url = {https://doi.org/10.1093/mnras/stz1466},
    eprint = {https://academic.oup.com/mnras/article-pdf/487/3/3112/43778821/stz1466.pdf},
}

@ARTICLE{sat3,
       author = {{van Uitert}, Edo and {Cacciato}, Marcello and {Hoekstra}, Henk and {Brouwer}, Margot and {Sif{\'o}n}, Crist{\'o}bal and {Viola}, Massimo and {Baldry}, Ivan and {Bland-Hawthorn}, Joss and {Brough}, Sarah and {Brown}, M.~J.~I. and {Choi}, Ami and {Driver}, Simon P. and {Erben}, Thomas and {Heymans}, Catherine and {Hildebrandt}, Hendrik and {Joachimi}, Benjamin and {Kuijken}, Konrad and {Liske}, Jochen and {Loveday}, Jon and {McFarland}, John and {Miller}, Lance and {Nakajima}, Reiko and {Peacock}, John and {Radovich}, Mario and {Robotham}, A.~S.~G. and {Schneider}, Peter and {Sikkema}, Gert and {Taylor}, Edward N. and {Verdoes Kleijn}, Gijs},
        title = "{The stellar-to-halo mass relation of GAMA galaxies from 100 deg$^{2}$ of KiDS weak lensing data}",
      journal = {MNRAS},
         year = 2016,
        month = jul,
       volume = {459},
       number = {3},
        pages = {3251-3270},
          doi = {10.1093/mnras/stw747},
archivePrefix = {arXiv},
       eprint = {1601.06791},
 primaryClass = {astro-ph.GA},
       adsurl = {https://ui.adsabs.harvard.edu/abs/2016MNRAS.459.3251V}
}

@article{sat4,
    author = {Wojtak, Radosław and Mamon, Gary A.},
    title = {Physical properties underlying observed kinematics of satellite galaxies},
    journal = {Monthly Notices of the Royal Astronomical Society},
    volume = {428},
    number = {3},
    pages = {2407-2417},
    year = {2013},
    month = {01},
    issn = {0035-8711},
    doi = {10.1093/mnras/sts203},
    url = {https://doi.org/10.1093/mnras/sts203},
    eprint = {https://academic.oup.com/mnras/article-pdf/428/3/2407/3692699/sts203.pdf},
}

@article{sat5,
    author = {More, Surhud and van den Bosch, Frank C. and Cacciato, Marcello and Skibba, Ramin and Mo, H. J. and Yang, Xiaohu},
    title = {Satellite kinematics – III. Halo masses of central galaxies in SDSS},
    journal = {Monthly Notices of the Royal Astronomical Society},
    volume = {410},
    number = {1},
    pages = {210-226},
    year = {2011},
    month = {01},
    issn = {0035-8711},
    doi = {10.1111/j.1365-2966.2010.17436.x},
    url = {https://doi.org/10.1111/j.1365-2966.2010.17436.x},
    eprint = {https://academic.oup.com/mnras/article-pdf/410/1/210/18443023/mnras0410-0210.pdf},
}

@article{sat6,
doi = {10.1088/0004-637X/690/2/1488},
url = {https://doi.org/10.1088/0004-637X/690/2/1488},
year = {2008},
month = {dec},
publisher = {The American Astronomical Society},
volume = {690},
number = {2},
pages = {1488},
author = {Klypin, Anatoly and Prada, Francisco},
title = {TESTING GRAVITY WITH MOTION OF SATELLITES AROUND GALAXIES: NEWTONIAN GRAVITY AGAINST MODIFIED NEWTONIAN DYNAMICS},
journal = {The Astrophysical Journal}
}

@ARTICLE{hernquist,
       author = {{Hernquist}, Lars},
        title = "{An Analytical Model for Spherical Galaxies and Bulges}",
      journal = {\apj},
         year = 1990,
        month = jun,
       volume = {356},
        pages = {359},
          doi = {10.1086/168845},
       adsurl = {https://ui.adsabs.harvard.edu/abs/1990ApJ...356..359H}
}

\end{document}